\documentclass[twocolumn,showpacs,floatfix,superscriptaddress,amsmath,amssymb,aps,prb]{revtex4-1}
\usepackage{psfrag}
\usepackage{wrapfig}
\usepackage{subfigure}
\usepackage{psfrag}
\usepackage{tabularx} 
\usepackage{color} 
\usepackage[dvips]{graphicx}
\DeclareGraphicsExtensions{.eps}

\begin{document}
\title{Vortex and disclination  structures in a nematic-superconductor state}
\author{Daniel G. Barci}
\author{Rafael V. Clarim}
\author{N.\  L.\   Silva J\'unior}
\affiliation{Departamento de F{\'\i}sica Te\'orica,
Universidade do Estado do Rio de Janeiro, Rua S\~ao Francisco Xavier 524, 20550-900,  Rio de Janeiro, RJ, Brazil.}
\date{\today}
\begin{abstract}
The nematic-superconductor state is  an example of a quantum liquid crystal that  breaks gauge as well as rotation invariance. It was conjectured to exist in the pseudogap regime of the cuprates high $T_c$ superconductors.
The nematic-superconductor state is characterized by two complex order parameters: one of them is related with superconductivity and the other one describes a nematic order.  It supports two main classes of topological defects: half-vortices and disclinations. In this paper we present a Ginzburg-Landau approach to study the structure of these topological defects.
Due to a geometrical coupling between the superconductor and the nematic order parameters, we show that vortices are strongly coupled with disclinations. 
We have found a restoring force between vortices and  disclinations that produces harmonic excitations whose natural frequency depends on the geometrical coupling constant and the superconductor condensation energy. Moreover, in a regime with high density of defects,  we have found a structural phase transition between vortex-disclination lattices with different symmetries. 
\end{abstract}
%74.20.De Phenomenological theories (two-fluid, Ginzburg-Landau, etc.) 
%74.25.Uv 	Vortex phases (includes vortex lattices, vortex liquids, and vortex glasses)
%75.70.Kw 	Domain structure (including magnetic bubbles and vortices) (for domain structure in ferroelectricity and antiferroelectricity, see 77.80.Dj)
%74.72.Kf   Pseudogap regime 
%74.25.Ha 	Magnetic properties including vortex structures and related phenomena (for vortices, magnetic bubbles, and magnetic domain structure, see 75.70.Kw)
%74.25.Dw 	Superconductivity phase diagrams
\maketitle

%%%%%%%%%%%%%%%%%%%
\section{Introduction}
\label{Sec:Introduction}
%%%%%%%%%%%%%%%%%%%

It is by now well stablished that electronic anisotropies play an important role in several 
strongly correlated systems\cite{FradKiv2010}. Strong repulsive interactions could spontaneously 
break  lattice translation as well as rotation invariance. Depending on the broken symmetries, different ordered phases  can be classified in a very similar way to liquid crystals. For instance, a stripe phase, that breaks translation invariance in one direction, has the same symmetries of the {\it smectic} phase. On the other hand, phases breaking only rotation invariance, but preserving translations, are called {\it nematic} phases.  This equivalence allows us to generically call these phases  ``quantum liquid crystals'' \cite{FrKi1999}. In particular,  theories  for  the quantum Hall smectic phase\cite{FrKi1999,BaFrKiOg2002} and the nematic Fermi   liquid\cite{OgKiFr2001,Lawler2006} was presented some time ago.   

In some strongly correlated systems, the superconductor (SC) order parameter can break  rotational as well as discrete translational invariance, in such a way that  the orientational and positional orders are intertwined with the SC and magnetic orders\cite{BeFrKiTr2009,FrKiTr2015}. In these cases, the traditional classification of s-wave, d-wave, etc., coming from the irreducible representation of the lattice point symmetry  group,  does no longer apply.
An important particular example is the {\em pair density wave}  state (PDW)\cite{berg-2007, berg-2008a,FrKiTr2015}. It was originally  proposed to describe the striking dynamical dimensional decoupling, observed in $La_{2-x}Ba_xCuO_4$ near $x=1/8$\cite{li-2007,tranquada-2008}.  Similar effects have been observed in stripe-ordered $La_{1.6-x}Nd_{0.4}Sr_xCuO_4$\cite{tajima-2001,ding-2008} and in the magnetic-field induced stripe-ordered phase of $La_{2-x}Sr_xCuO_4$\cite{schafgans-2008}. Roughly speaking, the PDW state can be thought as a condensate of Cooper pairs with  finite momentum. Interestingly, there is a  recent claim that  a PDW state has been measured in $Bi_2Sr_2CaCu_2O_{8+x}$ by means of  nanometre-resolution scanned Josephson tunneling microscopy \cite{Hamidian2016}.

The nematic superconductor, that we study in this paper,  is induced by the more basic PDW state.
Consider, for instance, the simplest unidirectional PDW  state characterized by a wave vector $q$,
\begin{equation}
\Delta_{PDW}=\Delta_q e^{i q x} +\Delta_{-q} e^{-i q x}\; ,
\end{equation} 
with $\Delta_q\neq \Delta_{-q}^*$. Both order parameters, $\Delta_q=|\Delta_q|\exp(i\theta_q)$ and $\Delta_{-q}=|\Delta_{-q}|\exp(i\theta_{-q})$,  represent Cooper pairs (charge $2e$) with momentum $q$. Thus, the state is characterized by two complex order parameters. The corresponding  phases can be identified as the superconducting phase
$\theta_+=(\theta_q+\theta_{-q})/2$ and the smectic displacement $\theta_-=(\theta_q-\theta_{-q})/2$. Since  
$\theta_{\pm q}$ are defined modulo $2\pi$, the related phases $\theta_{\pm}$ are defined modulo $\pi$.  
Thus, the PDW state supports several topological defects, such as vortices, double dislocations and half-vortices bounded to single dislocations\cite{berg-2009}.
By means of thermal melting, several phases can be reached from the PDW state,  producing a very rich phase diagram\cite{barcifradkin-2011}. One of these phases, called {\em nematic-superconductor} (NSC)  is reached from the PDW state by thermal proliferation of double dislocations. The proliferation of dislocations restores the translation invariance, retaining the SC as well as the orientational (nematic) order.
The NSC is understood as a quadratic combination $\psi\propto\Delta_q\Delta_{-q}$. In this way, $\psi$ is a complex homogeneous order parameter, since the combination of 
$\Delta_q$ and $\Delta_{-q}$  cancels any modulation in $q$. However, the state still ``remember'' the original orientation of  $q$, thus, producing an homogeneous anisotropic state. Interestingly, from its definition, it is simple to check that, under gauge transformations, it transforms with double the phase of the PDW state, since  $\psi\propto|\Delta_q\Delta_{-q}| \exp(2i\theta_+)$. Then, this superconductor state should be interpreted as a condensation of four particles and not of Cooper pairs. For this reason it is also called ``charge-$4e$ nematic superconductor'' ($4e$-NSC).
 
 Nematic fluctuations enhance this state since, for weak lattice couplings, the PDW phase turns out to be unstable. Indeed, in the limit where the lattice is completely decoupled, two-dimensional positional order cannot exist, due to linearly divergent fluctuations\cite{barcifradkin-2011}. It is highly hypothesized that the NSC state could exist in the pseudogap region of cuprates. Indeed, several experimental clues point in this direction. For instance,  fluctuating stripes have been measured\cite{Parker2010} at the onset of the pseudo-gap state of $Bi_2Sr_2CaCu_2O_{8+x}$.  Moreover, measurements of the Nernst effect in $YBa_2Cu_3 O_y$\cite{Daou2010} showed that the pseudogap temperature coincides with the appearance of a strong in-plane anisotropy of electronic origin, compatible with the electronic nematic phase\cite{Lawler2006}. 

The NSC state has extremely interesting properties. Its main topological defects are  half-vortices (vortices with half a flux)  and disclinations.  Thus, thermal melting could produce a metallic nematic phase (by proliferation of half-vortices) or even an isotropic superconductor (by proliferation of disclinations). However,  the scenario is not so simple since, as we will show, vortices and disclinations are strongly interacting.   

In this paper, we analyze the structure of the topological defects supported by the NSC state.  By means of a  Ginzburg-Landau  (GL) theory, we compute the vortex and the disclination profiles in two different regimes: a very diluted regime,  where vortices and disclinations can be considered isolated (with axial symmetry), and a high density regime, where vortices and disclinations tend to form lattice structures.  The effect of local nematicity is to produce a fluctuating metric which  produces a geometrical coupling between the nematic and  the SC order parameter. In some sense, the system  behaves similarly to an order parameter living in a curved surface\cite{kamien-2002}. 
We have found that the geometrical coupling tends to align the nematic director perpendicularly or parallel  to the supercurrent, depending on the sign of the coupling.  This effect implies that vortices are strongly tightened to disclinations. We will show that, at short distances, both topological defects interact through a quadratic potential. The excitations are harmonic oscillations with frequency $\Omega^2\sim \lambda_{SC} \Lambda$, where $\lambda_{SC}$ is the SC condensation energy and $\Lambda$ is the geometrical coupling constant. The potential remains attractive when the separation between the vortex and the disclination growths, having a logarithmic dependence at large distances.
 In the high density regime, this strongly attractive interaction induces the system to be arranged in a lattice of vortices tightly bounded to disclinations. While vortices prefer to form triangular lattices, disclinations have a tendency to form square lattices. Then, there is a competition produced by the geometrical coupling, opening the possibility of a structural phase transition between  lattices with different symmetries. 
 
 The NSC state has never been directly detected, however, there is strong evidence that its parent PDW state has a relevant role near the pseudogap regime and could be recently been observed\cite{Hamidian2016}. A clear signature of the existence of the NSC state should be a detection of a half-flux ($hc/4e$)  vortex,  possibly by means of a SQUID loop arrangement\cite{berg-2009}. Due to the strong interaction between vortices and disclinations, in this paper we open the interesting possibility of probing the NSC state by using mechanical probes\cite{Locquet-1998,Sonin-1996}.

Along the paper we present  details of the model and calculations that conduce to the  above described main results. The paper is organized as follows: in section \ref{Sec:OrderParameters} we review the superconductor and the nematic order parameters and we show how to build the Ginzburg-Landau theory for the NSC state.  In \S \ref{sec:London} we analyze the simplest  approximation, in which the relevant degrees of freedom are the SC and the nematic phases. Section  \ref{sec:VortexDisclination}
 is the main part of the paper. In \S\ref{sec:isolatedVortex} we compute the profile of an isolated vortex-disclination  with axial symmetry while in the subsection 
 \ref{subsec:lattice} we analyze the case of a high density of vortices and disclinations showing the competition between different lattice symmetries. 
Finally, we discuss our results in \S \ref{sec:discussion} and reserve three appendices to show  computational  details.

%%%%%%%%%%%%%%%%%%%%%%%%%%%%%%%%%%%%%%%%%%%
\section{Order parameters and Ginzburg-Landau theory of the Charge-$4e$ Nematic Superconductor}
\label{Sec:OrderParameters}
%%%%%%%%%%%%%%%%%%%%%%%%%%%%%%%%%%%%%%%%%%
The nematic superconductor is an example of a quantum liquid crystal\cite{kivelson-1998,radzihovsky-2008}. It is an homogeneous electronic state that breaks gauge as well as rotation invariance. Thus,  it is necessary to deal with two order parameters, one of them complex, related with superconductivity and the other one related to the orientational order\cite{OgKiFr2001}. 
With the aim of  making this paper self-contained, we briefly review in this section the Ginzburg-Landau theory for the NSC state\cite{BeFrKiTr2009}, paying special attention on  the geometrical coupling induced by nematicity\cite{barcifradkin-2011}.

The simplest superconductor order parameter is given by a scalar  complex function
\begin{equation}
\psi(\vec x)=\rho(\vec x) e^{i\theta(\vec x)}\; .
\label{eq:psi}
\end{equation}
Moreover, the two-dimensional nematic order parameter is represented by a second order, traceless symmetric tensor ${\bf N}$, whose components are given by  
\begin{equation}
N_{ij}= 2 S(\vec x) \left\{ n_i(\vec x) n_j(\vec x)-\frac{1}{2} \delta_{ij}\right\} \; , 
\label{eq:Nij}
\end{equation}
with $i,j=x,y$. $S(\vec x)$ is the modulus of the order parameter and the unit vector $\hat n=(\cos\alpha,\sin\alpha)$ is the director of the nematic order. $N_{ij}$ is a quadratic function of the director, making it invariant under $\pi$-rotations, $\hat n(\vec x)\to -\hat n(\vec x)$.  In two dimensions, ${\bf N}$ has two independent components, $N_{xx}$ and $N_{xy}$, that could be arranged in a complex function  $Q= N_{xx}+ i N_{xy}$, in such a way that,  
\begin{equation}
Q(\vec x)= S(\vec x) e^{i 2\alpha(\vec x)}\; .
\label{eq:Q}
\end{equation}
The complex representation of the nematic order parameter is only possible in  two dimensions. In three dimensions, it is necessary to go back to the tensor representation since, in this case, there are more degrees of freedom and more possibilities for the nematic structure such as uniaxial and bi-axial nematics\cite{deGPr1998}. 

Both order parameters $\psi(\vec x)$ and $Q(\vec x)$ are  formally very similar, in the sense that both are complex functions.  However, their transformation properties are very different: $\psi(x)$  transforms under an {\em internal} $U(1)$ gauge symmetry group and it is a scalar field under external global rotations, {\it i.e.}, if we rotate the coordinates system by an angle $\varphi$, $\vec x'=R_{\varphi}(\vec x)$, then  $\psi'(\vec x')=\psi(\vec x)$. On the other hand, the nematic order parameter does not transform under the internal gauge group, however,   $Q'(\vec x')=e^{i2\varphi}Q(\vec x)$ under   rotations, since it comes from a second rank tensor structure.  The factor $2$ in the exponential enforces the nematic symmetry. These properties are very important in order to correctly built up a gauge and rotational invariant free energy. 

For uniform configurations, the Landau expansion for both order parameters can be made as usual. Assuming that near the transition $|\psi|$ is small, and asking for rotational and  gauge invariance, with the additional requirement of  analyticity,  we have the quartic potential for the superconductor order parameter
\begin{equation}
V_{\rm SC}= a |\psi|^2+\frac{b}{2} |\psi|^4= a \rho^2+\frac{b}{2} \rho^4\;, 
\label{eq:VSC}
\end{equation}
where we assume $b>0 $. The metallic/superconductor phase transition is controlled by the sign of $a=\tilde a (T-T_{SC})$, where $\tilde a$ is a constant and $T_{SC}$ is the mean-field superconductor critical temperature.
For the nematic order parameter we have an equivalent expansion
\begin{eqnarray}
V_{\rm N}&=& \frac{t}{2} Tr({\bf N}^2) + \frac{u}{4} Tr({\bf N}^4) \nonumber \\
&=& t |Q|^2+ \frac{u}{2} |Q|^4=t S^2+ \frac{u}{2} S^4 \; .
\label{eq:VN}
\end{eqnarray}
Since $u>0$, the isotropic/nematic transition is controlled by $t=\tilde t (T-T_{N})$, where $\tilde t$ is a constant and $T_{N}$ is a mean-field nematic critical temperature. We assume that $T_{N}> T_{SC}$,  such that  a metallic nematic phase exists at temperatures $T_{SC}<T<T_{N}$. The expression of Eq. (\ref{eq:VN}) is typical of two-dimensional nematic where $Tr( {\bf N}^{2n+1})=0$. Conversely, in three dimensions, $Tr( {\bf N}^{3})\neq 0$, producing a first order phase transition. 

The simplest way to couple $Q$ and $\psi$  taking into account  phase symmetry and rotation invariance is  through the quartic potential 
\begin{equation}
V_{SCN}= \frac{v}{2} |\psi|^2 Tr({\bf N}^2)=v |\psi|^2|Q|^2=v \rho^2S^2\; , 
\label{eq:VSCN}
\end{equation}
where $v$ is a coupling constant. For weak coupling, $|v/ub|<<1$, the homogeneous Landau free energy is minimized by, 
\begin{eqnarray}
\rho_m^2&=&-\frac{a}{b}+ t\left(\frac{v}{ub}\right)+ O[(v/ub)^2]\; , 
\label{eq:rhom}  \\
S_{m}^2&=&-\frac{t}{u}+ a\left(\frac{v}{ub}\right)+ O[(v/ub)^2] \; .
\label{eq:Sm} 
\end{eqnarray}
Thus, if $v<0$, the presence of one phase strengths the presence of the other one. However, for  $v>0$, both phases are competing. 

Coupling the nematic order parameter with inhomogeneous superconductor configurations  is more subtle. Since the nematic order parameter is a second rank tensor it couples with the derivatives  of the superconductor order parameter. Indeed, the simplest coupling of this type is $\nabla_i\psi^* N_{i,j}\nabla_j\psi$, which is  obviously globally gauge invariant and rotational invariant. In this sense, the nematic order parameter behaves as a fluctuating metric\cite{barcifradkin-2011}. It is worth to mention that the effect of nematicity as an effective fluctuating metric was recently found in other electronic systems such as fractional quantum Hall systems\cite{You2016}.
We can built the GL free energy by considering that the SC order parameter lives in a curved space\cite{kamien-2002}, characterized by the metric
\begin{equation}
g_{ij}(\vec x)=\delta_{ij}+\frac{\Lambda}{S_m} N_{ij}(\vec x)\; , 
\label{eq:gij}
\end{equation}
where the constant $\Lambda$  measures the geometrical coupling and $S_m$ is just a normalization to get the coupling dimensionless. We note here that $\Lambda$ could be any small real number, positive or negative. On the other hand, remembering that the nematic tensor is invariant under 
rotations by $\pi$ and changes sign under rotations by $\pi/2$, a change of sign in  $\Lambda$ is equivalent to a global rotation of  $\pi/2$ of the director $\hat n(x)$. 

We will consider essentially three contributions to the GL free energy: the potential terms, given by Eqs. (\ref{eq:VSC}), (\ref{eq:VN}) and (\ref{eq:VSCN}) that involve only scalar homogeneous couplings; the derivative terms proportional to   $g^{ij}\nabla_i\psi^*\nabla_j\psi$ where the geometrical coupling plays an important role; and the inhomogeneous nematic terms that take into account the ``elastic'' properties of the nematic degrees of freedom.  Moreover, as usual, we need to minimally couple the electromagnetic field in order to force local gauge invariance. 
 In this way, the  GL free energy reads, 
\begin{eqnarray}
F_{LG}&=&\!\int \!\! d^2x   \sqrt{\det g}\left\{ \alpha_{s} g^{ij} (D_i \psi)^*(D_j \psi)+ \alpha_n\vec\nabla Q^*\cdot\vec\nabla Q\right.\nonumber \\
&+& \left. V_{SC}+V_{N}+V_{NSC}+ \frac{1}{8\pi} B^2\right\}\; .
\label{eq:FLG}
\end{eqnarray}
The first term in Eq. (\ref{eq:FLG}) codifies  the  interaction between derivatives of the SC order parameter. It contains couplings with nematic degrees of freedom through  the metric $g^{ij}$ given by Eq. (\ref{eq:gij}). The constant 
 $\alpha_s$ measures the superconducting stiffness and is related with the coherent length, as we will show in the next sections.  The second term, proportional to $\alpha_n$, is the simplest nematic elastic energy; we have  considered, for simplicity, that all Frank constants are equal\cite{deGPr1998}. 
The second line of Eq. (\ref{eq:FLG}) contains the potentials and the energy density of the magnetic field 
$B=\vec\nabla\times\vec A$.
The covariant derivatives are given by
\begin{equation}
D_i=\nabla_i -i 4e A_i\;.
\label{eq:D}
\end{equation}
The value of the electric charge of elementary excitations is not fixed in a GL theory. In usual superconductors we fix it to $2e$ since we associate the modulus of the order parameter with  the density of Cooper pairs.  In the case of the NSC, we have no microscopic theory (such as BCS) to guide us to fix the charge. Conversely, we fix it  to  $4e$, since we understand the NSC  as a melted PDW state as explained in  section \ref{Sec:Introduction}.
Finally, the integration measure $d^2x   \sqrt{\det g}$, is the usual invariant measure under re-parametrizations.

Eq. (\ref{eq:FLG}) is the main result of this section and is the starting point of the subsequent analysis of the topological defects structure.
This model has certain similarities with other superconductor states described by multicomponent  order parameters\cite{Col2005,MiloradPerali2015} . For instance, two-band superconductors with different coherent lengths   admit  vortices with fractional magnetic  flux.\cite{Babaev-2002}. Moreover, using a two component order parameter theory, it has been recently conjectured that a square lattice of  skyrmions could be topologically stable in the pseudogap regime\cite{Mauro2015}.  The essential difference with our model is that the nematic order parameter $Q$ does not couple with the vector potential $A_i$ in a minimal way, but it does couple through the metric. There is also a nontrivial geometrical coupling between both order parameters that uniquely characterize the NSC state.
In the next sections we analyze the influence of these couplings on the topological configurations that minimize the free energy, Eq. (\ref{eq:FLG}).

%%%%%%%%%%%%%%%%%%%%%%%%%%%%%%%%%%%%%%%%%%
\section{ \label{sec:London}
Warming up: London approximation}
%%%%%%%%%%%%%%%%%%%%%%%%%%%%%%%%%%%%%%%%%%%

Some general features of the  topological defects structure can be visualized using a simpler free energy, obtained in analogy with the London approximation in usual superconductors\cite{annett2004superconductivity}.
The  system described by Eq. (\ref{eq:FLG}) has three typical length scales. The superconducting coherent length
\begin{equation}
\xi_s=\sqrt{\frac{\alpha_s}{|a|}}
\end{equation} 
that relates the coefficient of the derivative of the order parameter with the curvature of the SC potential,   indicating the typical length scale of the modulations of the SC order parameter. There is an analogous lengh scale for the coherence of the nematic order parameter given by
\begin{equation}
\xi_n=\sqrt{\frac{\alpha_n}{|t|}}
\end{equation} 
Finally, the London penetration length $\lambda_L$ is related with the electromagnetic response of the system, and represents the typical length in which a magnetic field can penetrate a superconductor state. In our system, it is given by
\begin{equation}
\lambda_L=\sqrt{\frac{1}{138 \pi e^2 \alpha_s}\left(\frac{b}{a}\right)} +O(\Lambda).
\end{equation} 
The numerical coefficient is half the usual one, since the elementary charge of the NSC is $4e$. Moreover, it has small corrections due to the anisotropy of order $\Lambda$.
Thus, we can define two dimensionless constants given by 
\begin{equation}
\kappa_s=\frac{\Lambda_L}{\xi_s}\mbox{~~~and~~~}\kappa_n=\frac{\Lambda_L}{\xi_n}.
\label{eq:kappa}
\end{equation} 
The first one is the usual Abrikosov parameter while the second one is a similar parameter for the nematic component. The penetration length is the same for both definitions since the nematic order parameter does not couple in a minimal way with the electromagnetic field.
  
In a temperature  regime where $T<<T_{SC}<T_{N}$, we can ignore $\rho$ and $S$ fluctuations. 
From Eq. (\ref{eq:FLG}), and considering  $S=S_m$ and $\rho=\rho_m$ given by Eqs. (\ref{eq:rhom}) and 
(\ref{eq:Sm}), we find the following free energy for the superconducting phase $\theta(x)$ and the nematic orientation $\alpha(x)$, 
\begin{eqnarray}
F_{L}&=&\int d^2x\left\{  \rho_s\left| \vec\nabla\theta+4e \vec A \right|^2+K |\vec\nabla\alpha|^2 \right. \nonumber  \\
 &&+\left.\lambda\left(\hat n\cdot (\vec\nabla\theta+4e \vec A)\right)^2+\frac{1}{8\pi}\left(\vec\nabla\times\vec A\right)^2\right\} \; ,
\label{eq:FLondon}
\end{eqnarray} 
where 
 $\rho_s=\alpha_s\rho_m^2(1-\Lambda)$,  $K=4\alpha_n S^2_m$ and
 $\lambda=2\alpha_s\rho_m^2\Lambda$. This is a quite good  approximation for low temperatures and systems with   $\kappa_s>>1$ and 
$\kappa_n>>1$.

For weak external magnetic fields, the system is in a Meissner phase. It has phase coherence and the magnetic field is completely expelled from the sample.  However, near a critical value $H_{c_1}\sim \phi_0/\lambda_L^2$, determined by the penetration length $\lambda_L$ and the magnetic flux quantum  $\phi_0$,  the system can lower its energy by letting a quantized magnetic flux to penetrate the sample in a limited region determined by the coherent length 
$\xi_s$. Once the magnetic field penetrates the sample, circulating supercurrents  suppress the SC order parameter inside the core.  On the other hand, for large distances, the screening currents produce a magnetization that suppresses  the magnetic field on a length scale $\lambda_L$.  For longer distances the current density rapidly vanishes.  The nematic order parameter can present topological defects called disclinations\cite{ChLu1995,deGPr1998}. Disclinations are regions where the director $\hat n(x)$ has a discontinuity. In two dimensions, disclinations are points-like objects in such a way that $\oint_C d\vec\ell\cdot\hat n=2\pi$, provided  the discontinuity is inside the closed path $C$. These defects are composed by a core where the system is essentially  isotropic.   
At very low temperatures, much smaller than the isotropic-nematic transition we expect only few disclinations since the main mechanism to produce them are thermal fluctuations.  However, due to the geometrical coupling with the SC order parameter, the presence of vortices always induces disclinations. Thus, in our model, disclinations are indirectly driven by the magnetic field.  

\begin{figure}[hbt]
\begin{center}
\includegraphics[width=0.3\textwidth]{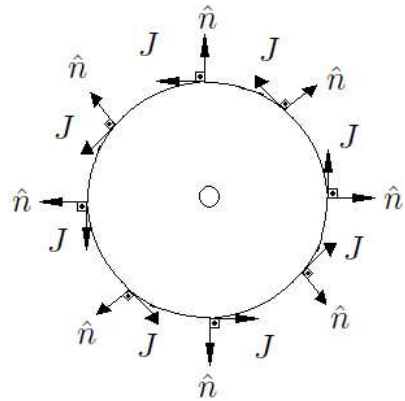}%}
\end{center}
\caption{Isolated vortex attached to a disclination. The radial director $\vec n(\vec x)$ is locally perpendicular to the vortex current $\vec J_{sc}(\vec x)$.  Inside the core, the system is  an isotropic metal.}
\label{fig:VortexDisclination}
\end{figure}

In the absence of the geometrical coupling $(\lambda=0)$, Eq. (\ref{eq:FLondon}) reduces to two decoupled $XY$ models\cite{berg-2009}. In this context, by rising temperature,  we expect two independent Kosterlitz-Thouless transitions; one of them driven by vortex unbinding producing a metallic nematic phase\cite{Lawler2006}  and the other one by unbinding disclinations, reaching a completely isotropic metallic phase.  
This scenario changes in the presence of the geometrical coupling since it forces the director to point perpendicular or parallel to the supercurrent depending on the sign of $\lambda$. This effect is easily seen by observing that the last term of Eq. (\ref{eq:FLondon}) is proportional to $\lambda(\hat n\cdot \vec J_{sc})^2$ with the supercurrent $\vec J_{sc}\sim \vec\nabla\theta+4e \vec A$. To minimize this term, $ \hat n \perp J_{sc}$ for $\lambda >0$. Converselly, for $\lambda<0$  the energy is minimized considering $ \hat n \parallel J_{sc}$. Thus, in either case, currents induce nematicity. In particular, the  presence of a vortex induces a disclination configuration, as shown in Fig.  (\ref{fig:VortexDisclination}) for the case $\lambda> 0$.

To be specific, let us  minimize the free energy by computing 
$\delta F/\delta \theta=0$ and $\delta F/\delta \alpha=0$. 
We obtain the following differential equations (for simplicity we put $\vec A=0$), 
\begin{eqnarray}
\nabla^2\theta&+&\frac{\lambda}{\rho_s}\left(\nabla_n^2\theta+ \nabla_n\theta\nabla_{n_\perp}\alpha
\right) =0 \; , 
\label{eq:nablatheta}\\
\nabla^2\alpha&-&\frac{\lambda}{K} \nabla_n\theta\nabla_{n_\perp}\theta =0 \; , 
\label{eq:nablaalpha}
\end{eqnarray}
where we have defined the following scalar differential operators
\begin{eqnarray}
\nabla_n&=&\hat n\cdot \vec \nabla  \; , \\
\nabla_{n_\perp}&=&\hat n\times \vec \nabla \; .
\end{eqnarray}
$\nabla_n$ and $\nabla_{n_\perp}$ are directional derivatives parallel and perpendicular to the director $\hat n(x)$, respectively.

We find three types of  configurations that solve  Eqs. (\ref{eq:nablatheta}) and (\ref{eq:nablaalpha}):
\begin{itemize}
\item[a)] The trivial solution is $\theta(x)= \theta_0$,  $\alpha(x)= \alpha_0$ where $\theta_0$ and $\alpha_0$ are two arbitrary constants. This solution corresponds to an anisotropic  superconductor state with global phase $\theta_0$ and the nematic director aligned with the direction $\hat n_0=(\cos\alpha_0,\sin\alpha_0)$.
\item[b)] Isolated disclinations for which $\theta(x)=\theta_0$, and for instance, $n_i(x)= x_i/r$ for $r\neq 0$.  This solution has 
zero supercurrent $\vec \nabla \theta=0$, and the director is in a radial topological configuration.  
\item[c)] A vortex attached to a disclination  in such a way that the director is perpendicular to the supercurrent at all points. One of these configurations is  $\nabla_i \theta=\epsilon_{ij} x_j/r^2$ and $n_i(x)= x_i/r$ for $r\neq 0$. $\epsilon_{ij}$ is the antisymmetric Levi-Civita tensor, thus, $\hat n(x)\cdot\vec\nabla\theta(x)=0$.  We depict this solution in Fig. (\ref{fig:VortexDisclination}).  
\end{itemize}
%%%%%%%%%%%%%%%%%%%%%%
%%%%%%%%%%%%%%%%%%%%%
Interestingly, an isolated vortex is not a solution of Eqs. (\ref{eq:nablatheta}) and (\ref{eq:nablaalpha}), since the geometrical coupling forces the director to be perpendicular (or parallel) to the current streamlines, producing a disclination.

Therefore, in the London approximation, the thermal melting of the NSC state can be produced in two ways:  by unbinding isolated disclinations, which restores isotropy but does not affect the SC, or by the proliferation of vortices tightly bounded to disclinations.  It is timely to notice that  this mechanism is proper of isotropic interactions. The coupling to  lattice anisotropy changes  this scenario since the nematic transition becomes Ising like and it is driven by the proliferation of domain walls. In this case, vortices are no longer bounded to disclinations.  

\begin{figure}[hbt]
\begin{center}
\includegraphics[width=0.52\textwidth]{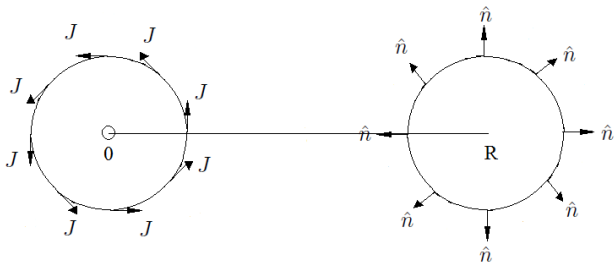}%}
\end{center}
\caption{Vortex and Disclination shifted by a distance $R$ in the $x$ direction}
\label{fig:VDR}
\end{figure}

The energy associated with a vortex-disclination configuration in the London approximation is simple to compute. For $\Lambda>0$, $J_{sc}\perp\hat n$ and  the energy has essentially two independent contributions $F_{vd}=F_v+F_d$. The energy of the vortex is approximately, $F_v\sim (\phi_0^2/\lambda_L^2)\ln(\kappa_s)$. Since the disclination does not couple with the electromagnetic field, $E_d=2\pi K\ln(L/\xi_n)$, where $L$ is the linear size of the sample\cite{ChLu1995}. This infrared divergence could be cut off at the inter-disclination distance in the diluted regime. In order to  understand more deeply the interaction between vortices and disclinations, let us compute the energy needed to create a vortex-disclination pair separated by  a distance $R$.
Consider, for instance, the configuration depicted in Fig (\ref{fig:VDR}).
%%%%%%%%%%%%%%%%%%%%
%%%%%%%%%%%%%%%%%%%%
The vortex is centered at the origin, while the disclination is centered at a  distance $R$ along the $x$ axes,
\begin{eqnarray}
\nabla_i\theta&=& \epsilon_{ij} \frac{x_j}{x^2+y^2} \; ,
\label{eq:vortex}\\
n_i&=& \frac{x_i-R_i}{\sqrt{(x-R)^2+y^2}}\; ,
\label{eq:disclination}
\end{eqnarray}
with $R_x=R$ and $R_y=0$. 
We compute the energy difference $\Delta F(R)=F_{vd}(R)-F_{vd}(0)$ 
using Eq. (\ref{eq:FLondon}),
\begin{equation}
\Delta F(R)=\lambda\int d^2 x\; \left(\mathbf{n}(x)\cdot \mathbf{\nabla}\theta(x,R)\right)^2 \; .
\label{eq:DeltaF}
\end{equation}
Replacing Eqs. (\ref{eq:vortex}) and (\ref{eq:disclination}) into Eq. (\ref{eq:DeltaF}), and performing the integrals (see appendix \ref{ap:VD}) we find, 
\begin{equation}
\Delta F(R)=\pi |\lambda|\ln\left(\frac{R}{a}\right) \; ,
\label{eq:Deltalog}
\end{equation}
where $a$ is the vortex core and  $R>>a$.
Thus, at large distances,  vortices and disclinations have an {\em attractive} logarithmic  interaction, whose sign is independent of the sign of their topological charges.

%%%%%%%%%%%%%%%%%%%%%%%%%%%%%%%
\section{Vortex and disclination profiles}
\label{sec:VortexDisclination}
%%%%%%%%%%%%%%%%%%%%%%%%%%%%%%%
In this section we  study more closely the interplay between vortices and disclinations by analyzing the complete GL free energy. 
It is useful to re-write  Eq. (\ref{eq:FLG}), in terms of modulus and phases of both order parameters. We have 
\begin{equation}
F=F_{SC}+F_{N}+F_{NSC}\; ,
\label{eq:F}
\end{equation}
where
\begin{eqnarray}
F_{SC}&=&\!\! \int  \! d^2x \!  \left\{ \alpha_{s}\!\!\left( |\vec D \rho|^2\!\!+\! \rho^2|\vec\nabla \theta+4e\vec A|^2\right)\!+\!V_{SC}\!\right. \nonumber \\
&+&\left.\frac{1}{8\pi} \left(\vec\nabla\times\vec A\right)^2\right\} \; , \\
F_{N}&=&   \int d^2x   \left\{ \alpha_{n}\left( |\vec\nabla S|^2+ 4 S^2|\vec\nabla \alpha|^2\right)+V_{N}\right\}\; , 
 \\
F_{NSC}&=&\int d^2x     \times \nonumber \\
&\times& \left\{ 2\alpha_{s}\Lambda \frac{S}{S_m} \left[\left( |\hat n\cdot\vec D \rho|^2+ \rho^2[\hat n\cdot(\vec\nabla \theta+4e\vec A)]^2\right)\right.\right. \nonumber \\
&&-\left.\left. \frac{1}{2}\left( |\vec D \rho|^2+ \rho^2|\vec\nabla \theta+4e\vec A|^2\right)\right]
+ V_{NSC}\right\}. 
\label{eq:geometric}
\end{eqnarray}
The first and the second equations are the superconductor and nematic free energies respectively, while the last one describes the  interaction between the order parameters. While the first term of Eq. (\ref{eq:geometric}) describes the geometrical interaction, the last term is the potential given by Eq. (\ref{eq:VSCN}). The main purpose of this section is to understand the effect of these interactions on the vortex and disclination profiles. 
To do this we consider two different regimes.  For weak magnetic fields, near the critical value  $H\gtrsim H_{c1}$,  vortices are extremely diluted and  we can consider the case of an isolated vortex-disclination configuration. On the other hand, for higher magnetic fields,  near $H\lesssim H_{c2}$, there is high density of vortices and  we study the formation of vortices/disclinations  lattices  with different symmetries.  

%%%%%%%%%%%%%%%%%%%%%%%%%%%%%
\subsection{ \label{sec:isolatedVortex} Effect of the geometrical coupling in a single vortex-disclination profile}
%%%%%%%%%%%%%%%%%%%%%%%%%%%%%
Guided by the results obtained in the London approximation,  we look for a single vortex  solution attached to a disclination centered at the origin. For $\Lambda>0$, the director should be perpendicular to the current, then, the simplest vortex-disclination configuration with axial symmetry  can be written as
\begin{eqnarray}
\psi(\vec x)&=&\rho(r)\;e^{i\varphi} \;  , 
\label{eq:Vpsi} \\
Q(\vec x)&=& S(r)\;e^{i 2\varphi} \; , 
\label{eq:DQ}
\end{eqnarray}
where $(r,\varphi)$ are usual polar coordinates. The factor  $2$ in the exponential of Eq. (\ref{eq:DQ}) guarantees the nematic symmetry $\varphi\to\varphi+\pi$. With this configuration, the current is locally perpendicular to the disclination as depicted in Fig. (\ref{fig:VortexDisclination}). Conversely, if  $\Lambda<0$, the Ansatz is replaced   by changing $Q\to -Q$, or equivalently  $\varphi\to\varphi+\pi/2$, in Eq. (\ref{eq:DQ}). In this way, the director is parallel to the supercurrent. 
Replacing Eqs. (\ref{eq:Vpsi}) and (\ref{eq:DQ})  into (\ref{eq:F}), and minimizing the free energy with respect to the radial functions  $\rho(r)$ and $S(r)$,  we find the following set of coupled differential equations:
\begin{eqnarray}
\lefteqn{
\alpha_s\left[-\rho''-\frac{\rho'}{r} +\frac{\rho}{r^2}\right]+a\rho + b\rho^3} 
\label{eq:rhodiff} \\
&-&\frac{\alpha_s\Lambda}{S_m}
\left[S\left(\rho''+\frac{\rho'}{r}+\frac{\rho}{r^2}\right)+S'\rho'\right]+v \rho S^2=0\; , 
\nonumber \\ 
&\alpha_n&\left[-S''-\frac{S'}{r} +4\frac{S}{r^2}\right]+ t S + u S^3
\label{eq:Sdiff} \\
&+&\frac{\alpha_s\Lambda}{2 S_m}
\left[(\rho')^2-\frac{\rho^2}{r^2}\right]+v \rho^2 S=0    \; , 
\nonumber 
\end{eqnarray}
where the prime means total derivative with respect to $r$, {\em i.e.} $\rho'=d\rho/dr$, 
$\rho''=d^2\rho/dr^2$ and so on.  To obtain these equations, we have considered that the magnetic field is essentially constant up to $\lambda_L>>\xi_s$,  meaning that we are deep inside the type II superconductor regime\cite{degennes-1999}. We have essentially disregarded screening effects in such a way that  the solution is strictly reliable  for $0<r\lesssim\lambda_L$. However, we expect that at $r\sim \lambda_L$ the vortex  profile is already saturated to the value $\rho(r)\sim \rho_m$, since the typical modulation length is $0\leq r\lesssim \xi_s\ll \lambda_L$.   The boundary conditions are 
$\lim_{r\to 0}\rho=0$, $\lim_{r\to 0}S=0$, $\lim_{r\to \infty}\rho=\rho_m$ and $\lim_{r\to \infty}S=S_m$.  $\rho_m$ and $S_m$ are given by Eqs. (\ref{eq:rhom}) and (\ref{eq:Sm}) respectively. 
The first line of Eq. (\ref{eq:rhodiff}) is the vortex differential equation with axial symmetry, while the first line of Eq. (\ref{eq:Sdiff})  is the analogous equation for the disclination. On the other hand, the second line of Eqs. (\ref{eq:rhodiff}) and (\ref{eq:Sdiff}) contain the two main couplings: 
the geometrical one, proportional to $\Lambda$ and the mixed potential energy, proportional to $v$. 
Interestingly, the system is invariant under the transformation $\Lambda\to -\Lambda$ and  $S \to -S$.  Changing the sign of $S$ means to rotate the director in $\pi/2$. Thus, the vortex and disclination profiles  are insensitive to whether the disclination is formed by directors parallel or perpendicular to the supercurrents. 

It is useful to rewrite Eqs. (\ref{eq:rhodiff}) and (\ref{eq:Sdiff}) in dimensionless form.    
For this, we first introduce the functions $f(r)$ and $g(r)$,
\begin{equation}
\rho(r)=\rho_m\; f(r)\mbox{~~~~ and ~~~~} S(r)=S_m\; g(r) \; , 
\label{eq:fg}
\end{equation} 
in such a way that the boundary conditions now read, 
 $f(0)=g(0)=0$, $\lim_{r\to\infty} f(r)=1$ and  $\lim_{r\to\infty} g(r)=1$.
 Replacing Eq. (\ref{eq:fg}) into Eqs. (\ref{eq:rhodiff}) and (\ref{eq:Sdiff}), using Eqs. (\ref{eq:rhom})
and (\ref{eq:Sm}) and keeping just linear terms in $v$, we finally find, 
\begin{eqnarray}
\lefteqn{
\frac{1}{2\kappa_s^2}\left[-f''-\frac{f'}{r} +\frac{f}{r^2}\right]-(f-f^3) +v_1 f(g^2-f^2)} 
\nonumber \\
&-&\frac{ \Lambda}{2\kappa_s^2}
\left[g\left(f''+\frac{f'}{r}+\frac{f}{r^2}\right)+g'f'\right]=0 \; , 
 \label{eq:fdiff} \\
&&\!\!\!\!\!\!\!\!\frac{1}{2\kappa_n^2}\left[-g''-\frac{g'}{r} +4\frac{g}{r^2}\right]-(g-g^3)+v_2 g(f^2-g^2)
\nonumber  \\
&+&\frac{\Lambda}{4\kappa_s^2}\frac{\rho_m^2}{S_m^2}
\left[(f')^2-\frac{f^2}{r^2}\right]=0 \;.   
\label{eq:gdiff}
\end{eqnarray}
We chose to measure distances in units of $\sqrt{2}\lambda_L$. 
 $\kappa_s$ and $\kappa_n$ are given in Eq. (\ref{eq:kappa}). We have also introduced the couplings $v_1= (t/au) v$ and $v_2=(a/tb) v$, both of them proportional to $v$.

We are interested in the solutions of Eqs. (\ref{eq:fdiff}) and  (\ref{eq:gdiff}) paying special attention on the effect of the geometrical coupling on the vortex-disclination profile. 
Of course, there is no exact analytical solution to these equations. Thus,  we will analyze the behavior of $f(r)$ and $g(r)$  at two extreme limits , $r\to 0$ and $r\to \infty$. Then, we  propose a systematic variational approach to interpolate between these regions.

Very near the origin ($r<<1$), we expect a linear behavior for the vortex solution, $f(r)\sim r$.  On the other hand, due to nematic symmetry,  the disclination approaches zero quadratically  as $r\to 0$, $g(r)\sim r^2$. Then, we look for a solution in a  power series of the form, 
\begin{eqnarray}
f(r)&=&\left(\frac{r}{{\cal R}_v}\right)\left\{1+c_1 r^2+c_2 r^4+\ldots \right\} \; , 
\label{eq:rhoseries} \\
g(r)&=&\left(\frac{r}{{\cal R}_d}\right)^2\left\{1+d_1 r^2+d_2 r^4+\ldots\right\} \; .
\label{eq:Sseries}
\end{eqnarray} 
${\cal R}_v$ and ${\cal R}_d$ are related with the core extension of the vortex and the disclination respectively.
Replacing these expressions into Eqs. (\ref{eq:fdiff}) and  (\ref{eq:gdiff}), it is possible to compute the set of coefficients $\{ c_1,c_2,\ldots\}$ and $\{d_1,d_2,\ldots\}$  recursively.  
The leading order correction is (for simplicity we ignored the potential interaction $v$)
\begin{eqnarray}
c_1&=&-\frac{1}{4}\kappa_s^2  \; , \\
d_1&=&-\frac{1}{6}\kappa_n^2 \left(1+\frac{\Lambda}{4}\left(\frac{\kappa_s}{\kappa_n}\right)^2 \left(\frac{\rho_m}{S_m}\right)^2  \right) \; .  
\end{eqnarray}
We see that  $c_1$ is not affected by the 
geometrical coupling, while $d_1$ has a small correction, since $\kappa_s/\kappa_n$ and $\rho_m/S_m$ are order one, and  $\Lambda <<1$ .
 Even though the complete sets $\{ c_1,c_2,\ldots\}$ and $\{d_1,d_2,\ldots\}$  can be univocally determined by Eqs. (\ref{eq:fdiff}) and (\ref{eq:gdiff}), the leading order coefficients ${\cal R}_v$ and ${\cal R}_d$   remain arbitrary and cannot be determined by a short distance expansion.  These quantities can only  be  fixed by the behavior of the solutions at large distances. For this reason we need to  analyze the asymptotic behavior of the solutions. 
   
For  $r>>1/\kappa_s$ and $r>>1/\kappa_n$  we have, 
\begin{eqnarray}
f(r)&=& 1+ f_1(r) \; ,
\label{eq:f1}   \\
g(r)&=& 1+ g_1(r) \; ,
\label{eq:g1}
\end{eqnarray}
where $\lim_{r\to \infty} f_1(r)=0$ and  $\lim_{r\to \infty} g_1(r)=0$.
Introducing Eqs. (\ref{eq:f1}) and (\ref{eq:g1}) into Eqs. (\ref{eq:fdiff}) and (\ref{eq:gdiff}), and linearizing the equations at large $r$ we find
\begin{eqnarray}
(1-v_1) f_1+v_1 g_1 &=& -\left(\frac{1-\Lambda}{4\kappa_s^2}\right)\frac{1}{r^2}  \; , 
\label{eq:rho1diff} \\
v_2 f_1+(1-v_2)  g_1 &=& -\left(\frac{1-\frac{1}{8}\frac{\rho_m^2}{S_m^2}\Lambda}{\kappa_n^2}\right)\frac{1}{r^2} \; .
\label{eq:S1diff} 
\end{eqnarray}
We immediately find the asymptotic  solutions    
\begin{eqnarray}
f(r)&=&1- \frac{1}{4\kappa_f^2 r^2}  \;,
\label{eq:flongr} \\
g(r)&=&1- \frac{1}{\kappa_g^2 r^2}  \;,
\label{eq:glongr}
\end{eqnarray}
where at linear order in $v$,  
\begin{eqnarray}
\kappa_f^2&=&\frac{\kappa_s^2}{1-\Lambda}
\left\{1-v_1\left(1-4\frac{\kappa_s}{\kappa_n}\left[\frac{1-\frac{1}{8}\frac{\rho_m^2}{S_m^2}\Lambda}{1-\Lambda}\right]  \right)  \right\} \; , 
\label{eq:kappaf} \\
\kappa_g^2&=&\frac{\kappa_n^2}{1-\frac{1}{8}\frac{\rho_m^2}{S_m^2}\Lambda}
\left\{1-v_2\left(1-\frac{1}{4}\frac{\kappa_n}{\kappa_s}\left[\frac{1-\Lambda}{1-\frac{1}{8}\frac{\rho_m^2}{S_m^2}\Lambda}\right]  \right)  \right\}. \nonumber \\
& &  \label{eq:kappag}
\end{eqnarray}
The asymptotic behavior, Eqs. (\ref{eq:flongr}) and (\ref{eq:glongr}), resembles the usual isolated Abrikosov vortex\cite{Abrikosov-1957}, with renormalized parameters $\kappa$. In fact, taking  the limits $\Lambda\to 0$ and $v\to 0$, we should recover Abrikosov result for $f(r)$. As can be seen from Eq. (\ref{eq:kappaf}), $\lim_{\Lambda\to 0, v\to 0} \kappa_f=\kappa_s$.

\begin{figure}[hbt]
\begin{center}
\subfigure[The continuous line is the vortex profile $f(r)$ while the dashed line represents the disclination profile $g(r)$.  $\Lambda=0.4$. ]{\label{fig:profilesa}
\includegraphics[width=0.4\textwidth]{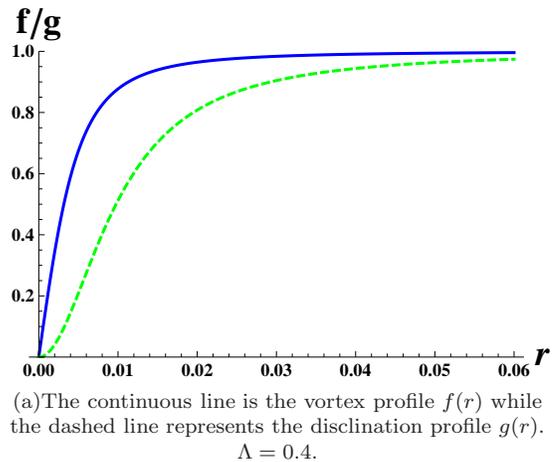}}
\subfigure[Vortex profile for different values of the geometrical coupling $\Lambda$. Dot-dashed, dashed and continuous  lines corresponds with $\Lambda=0, 0.4,0.8$, respectively. ]{\label{fig:profilesb}
\includegraphics[width=0.4\textwidth]{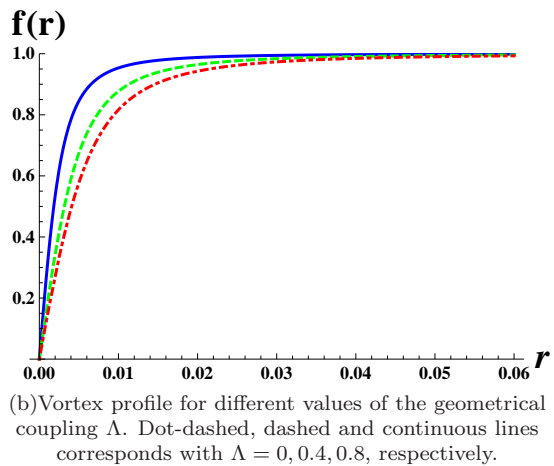}}
\end{center}
\caption{Vortex and Disclination profiles. We fixed $\kappa_s=\kappa_n=100$ and $v=0$ for all the curves. $r$ is measured in units of $\sqrt{2}\lambda_L$}
\label{fig:profiles}
\end{figure}

In order to completely determine the solutions, fixing the arbitrary constants ${\cal R}_v$ and  ${\cal R}_d$, it is necessary to interpolate between the short and large distance regions. A very simple variational Ansatz for the vortex profile is\cite{Schmid1966,Clem1975}
\begin{equation}
f(r)= \frac{\sqrt{2}\kappa_f r }{\sqrt{1+2\kappa_f^2 r^2}}\; ,   
\label{eq:vortex-ansatz}
\end{equation} 
while for the disclination the equivalent Ansatz read, 
\begin{equation}
g(r)= \frac{\kappa_g^2 r^2 }{1+\kappa_g^2 r^2}   \;.
\label{eq:disclination-ansatz}
\end{equation} 
Eqs. (\ref{eq:vortex-ansatz}) and (\ref{eq:disclination-ansatz}) correctly reproduce the assymptotic solutions Eqs. (\ref{eq:flongr}) and (\ref{eq:glongr}) for $\kappa_f r>>1$ and $\kappa_g r>>1$, while for small $r$, they reproduce the same power series structure than Eqs. 
(\ref{eq:rhoseries}) and (\ref{eq:Sseries}) respectively, {\it i.\ e.\  } ,  an odd power series for $f$ and an even power series for $g$.
We can improve the Ansatz in order to fit and arbitrary number of terms in the  small $r$  expansion. 
We present a systematic approach in appendix \ref{ap:Ansatz}. Fortunately, the leading order approximation, given by Eqs. (\ref{eq:vortex-ansatz}) and (\ref{eq:disclination-ansatz}), captures the main contribution of the geometrical coupling and is sufficient for the purpose of this section.
The leading order estimation of the vortex radius is
\begin{equation}
{\cal R}_v=\sqrt{\frac{1-\Lambda}{2}} \frac{1}{\kappa_s} \; .
\label{eq:Rv}
\end{equation}
On the other hand, at the same level of approximation, the disclination radius is 
\begin{equation}
{\cal R}_d=\sqrt{1-\frac{1}{8}\frac{\rho^2_m}{S^2_m}\Lambda}\;\frac{1}{\kappa_n} \; .
\end{equation}
In dimension-full quantities,  $R_v\sim \xi_s$. 
For most cuprates superconductors\cite{Lynn-Book-1990} $\xi_n\sim 20$ \AA  while the penetration length is approximately $\lambda_L\sim 2000$ \AA. Thus, $\kappa_n\sim 100 $ lies  deep in the type II region, where the approximations we used are accurate. On the other hand, the disclination radius $R_d\sim\xi_n$. An actual estimation of this length is  more speculative since we have no experimental inputs for $\alpha_n$. In our calculations, we assumed that $\lambda_L$ is the biggest length scale, in such a way that  $\kappa_s$, as well as, $\kappa_n$ are much bigger than one.  
In Fig.  (\ref{fig:profilesa}) we depict a typical vortex and disclination profile for $\kappa_s=\kappa_n=100$, $\Lambda=0.4$ and $\rho_m=S_m$. In Fig. (\ref{fig:profilesb}) we show the vortex profile for different values of the geometrical coupling $\Lambda$, displaying the radius dependence as given by Eq. (\ref{eq:Rv}).
 %%%%%%%%%%%%%%%%%%%%
%%%%%%%%%%%%%%%%%%%%
We see that the main effect of the geometrical coupling is to increase $\kappa_s$, or equivalently to decrease the superconductor coherence length. This is the same effect that disorder produce in a SC. In general, scattering from impurities reduces the value of 
$\alpha_s$ and, as a consequence, produces a smaller coherence length, $\xi_s$.  Then, just observing the vortex core extension, or the coherence length, is not sufficient to characterize the NSC state. 

In order to look for a clear signature, proper to the NSC state, let us compute the energy of a vortex and a disclination  shifted by  a distance $R$.   The energy difference, $\Delta F(R)$, between the shifted and coincident  vortex-disclination profile is given by
\begin{eqnarray}
\Delta F(R) &=&2\frac{\alpha_s\Lambda}{S_m} \int d^2x    S(r) \rho^2(r) \left(\hat n\cdot \vec\nabla\theta\right)^2  \\
&=& 2\alpha_s\Lambda\rho_m^2 R^2\int_0^\infty \frac{dr}{r} f^2(r)\int_0^{2\pi}d\varphi \frac{g(\bar r)}{\bar r^2}\sin^2\varphi  \; , \nonumber 
\end{eqnarray}
where $\bar r=r^2+R^2-2rR\cos\varphi$.
The integrals can be easily done in two limits,  $R<<R_v$ and $R>>R_v$. We find, 
\begin{equation}
\Delta F(R)\sim\left\{
\begin{array}{lll}
\frac{1}{2}\Omega^2 R^2    & & R<<{\cal R}_v  \\
 & & \\
\frac{\Omega^2}{\kappa_s^2}\ln(R/{\cal R}_v)   & &   R>> {\cal R}_v
\end{array}
\right.  \; ,
\label{eq:frequency}
\end{equation}
where the frequency  $\Omega^2= \lambda_{SC}  \Lambda$ and  $\lambda_{SC}=a^2/2b$ is the superconductor condensation energy. The estimation of the frequency $\Omega^2$ is equivalent to determine the condensation energy of a given material. This quantity is in general quite difficult to estimate. It could be obtained, for instance, form specific heat measurements\cite{Loram-1994,Loram-2001,Legget-2002}. 

Thus, the geometrical coupling produces an attractive force between the vortex and the disclination that does not depend on the sign of the topological charges. For large distances, the force is of the Coulomb type ($\sim 1/R$),  consistent with the London approximation. On the other hand, at  short distances, there is a linear restoring force producing oscillations  whose characteristic frequency is proportional to the condensation energy and the geometrical  coupling constant.  

%%%%%%%%%%%%%%%%%%%%%%%%%%%%%%
\subsection{\label{subsec:lattice}Vortex-Disclination lattices}
%%%%%%%%%%%%%%%%%%%%%%%%%%%%%%
In  stronger magnetic fields, a high density of vortices is present. We also expect a high density of disclinations, since in our model they  are  strongly tighten to vortices.  Since disclinations behave differently form vortices when coupled with electromagnetic fields, we expect unusual  properties of the vortex-disclination structure. As we will show, as a result of the geometrical coupling, a structural phase transition may take place between different lattice symmetries. In a different context, similar structural phase transitions were also predicted in lattices of half-vortices\cite{Chung-2010}.  

To explore this state of matter we closely follow  Abrikosov reasoning\cite{Abrikosov-1957}. We consider a magnetic field $H$ very near  $H_{c2}=\phi_0/4\pi\xi_s^2$, where $\phi_0$ is the quantum of magnetic flux, and $\xi_s=\sqrt{\alpha_s/a}$ is the superconductor coherent length. In this regime $\rho$ is very small, since we are near the metal-superconductor transition. Then, we can keep only quadratic terms in the free energy Eq. (\ref{eq:FLG}) in such a way that  the superconductor and nematic order parameters are essentially decoupled. 
Thus, in a quite good approximation, the magnetic field can be considered constant $H\sim H_{c2}$ and the linearized equations are essentially degenerated harmonic oscillators.  An approximate family of solutions can be cast in a linear superposition of solutions of the linearized equation in the form, 
\begin{equation}
\psi(x,y)=\sum_n C_n  e^{i \frac{2\pi}{b} n y} \exp\left\{-\frac{1}{2\xi_s^2} \left( x-\frac{2\pi \xi_s^2}{b} n\right)^2 \right\}\; , 
\end{equation}
where $b$ is the periodicity in the $y$ axes.  To impose periodicity in the $x$ axes, it is necessary to put constraints in the coefficients $C_n$.  For instance, for tetragonal symmetry,  $C_n=C_0$ for all $n$. On the other hand, for a triangular geometry $C_{n+2}=C_n$ for all $n$, and $C_0=i C_1$.

Then, we propose the variational vortex lattice solution as 
\begin{equation}
\psi(x,y)=C_1\; \chi(x,y) \; , 
\end{equation}
where $C_1$ is a variational parameter and $\chi(x,y)$ has different expressions depending on the symmetry. For tetragonal symmetry we have
\begin{equation}
\chi_{\Box}(x,y)=\sum_n  e^{i \frac{2\pi}{a} n y} \exp\left\{-\frac{1}{2\xi_s^2} \left( x-na \right)^2 \right\}\; , 
\label{eq:chibox}
\end{equation}
while for triangular symmetry
\begin{eqnarray}
\lefteqn{
\chi_{\triangle}(x,y)= \sum_n  e^{i \frac{4\pi}{\sqrt{3}a} n y} \left[
\exp\left\{-\frac{1}{2\xi_s^2} \left( x-na \right)^2\right\} \right. }   \nonumber \\
&+&\left. i e^{i \frac{2\pi}{\sqrt{3}a} n y} 
\exp\left\{-\frac{1}{2\xi_s^2} \left( x-(n+1/2)a \right)^2\right\}
\right] \; .
\label{eq:chitriangle}
\end{eqnarray}
In both cases, $a$ is the lattice constant. 

\begin{figure}[hbt]
\begin{center}
\subfigure[Contours of constant SC order parameter, $|\chi_\Box|^2=\mbox{constant}$ ]{\label{fig:VDSquare-a}
\includegraphics[width=0.33\textwidth]{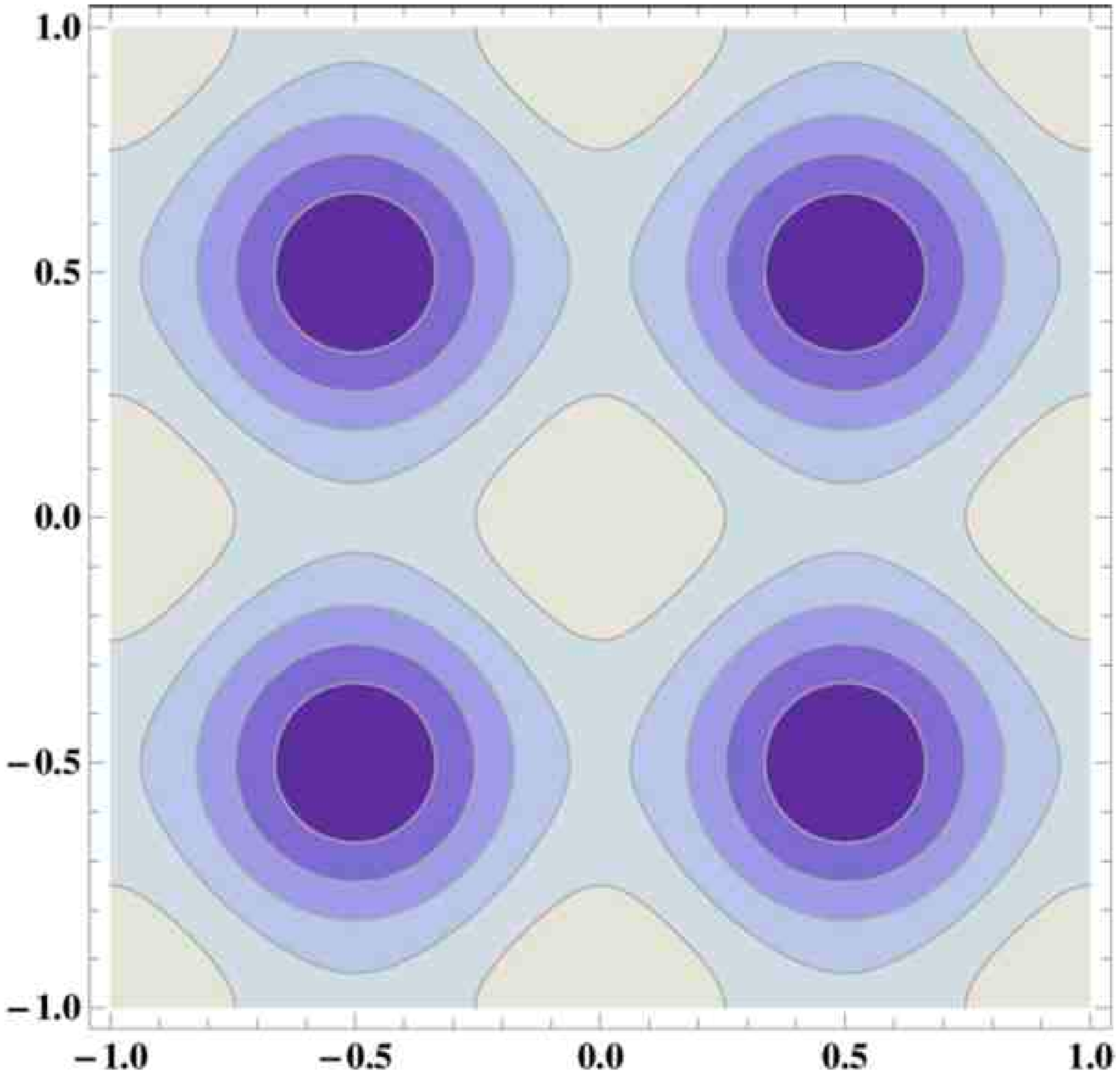}}
\subfigure[Contours of constant nematic order parameter, $|\vec\nabla|\chi_\Box|^2|=\mbox{constant}$ ]{\label{fig:VDSquarex-b}
\includegraphics[width=0.33\textwidth]{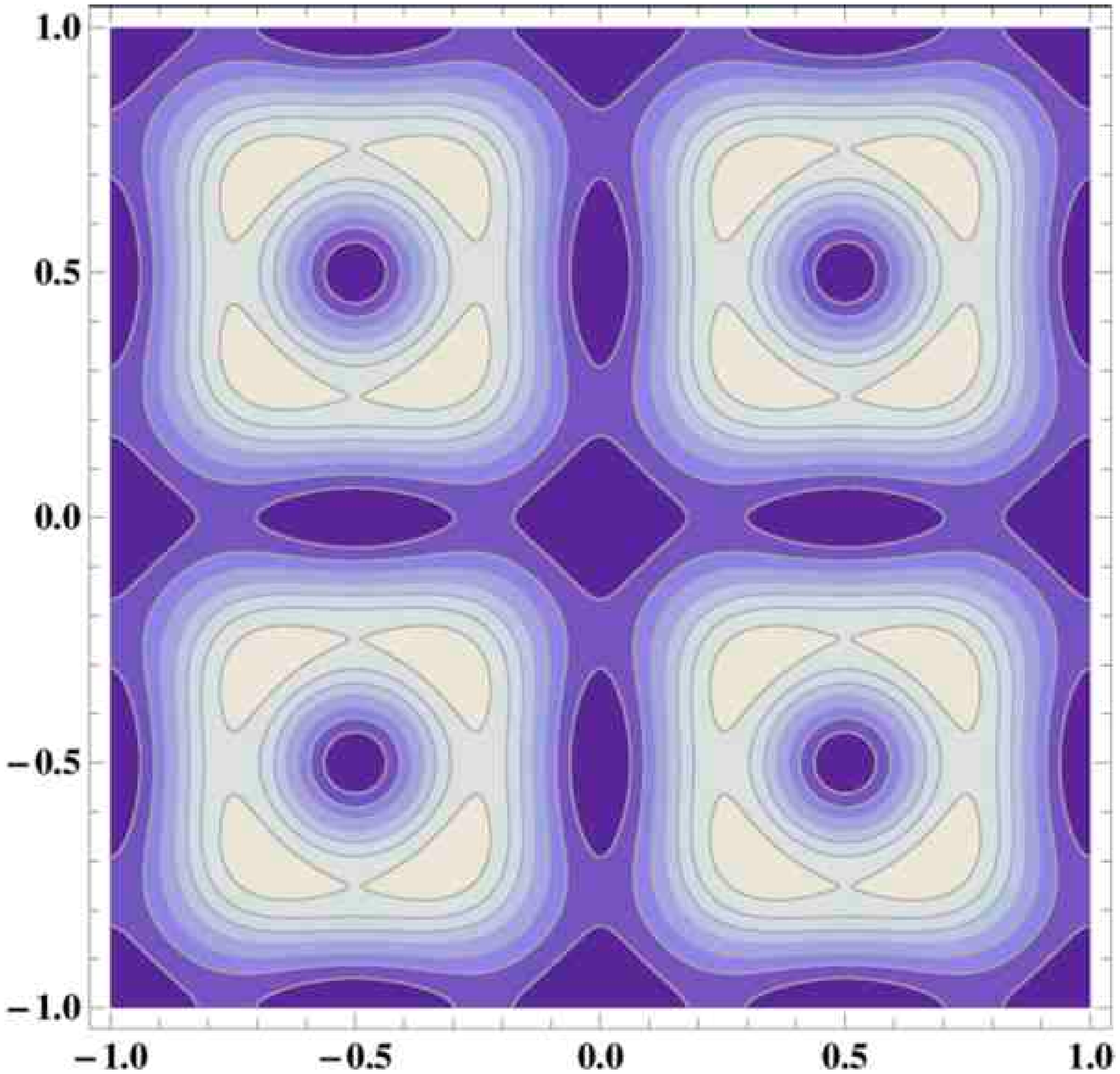}}
\end{center}
\caption{Vortex and Disclination lattice in a tetragonal configuration.}
\label{fig:VDSquare}
\end{figure}

\begin{figure}[hbt]
\begin{center}
\subfigure[Currents in a square vortex-disclination lattice]{\label{fig:currents-square-a}
\includegraphics[width=0.33\textwidth]{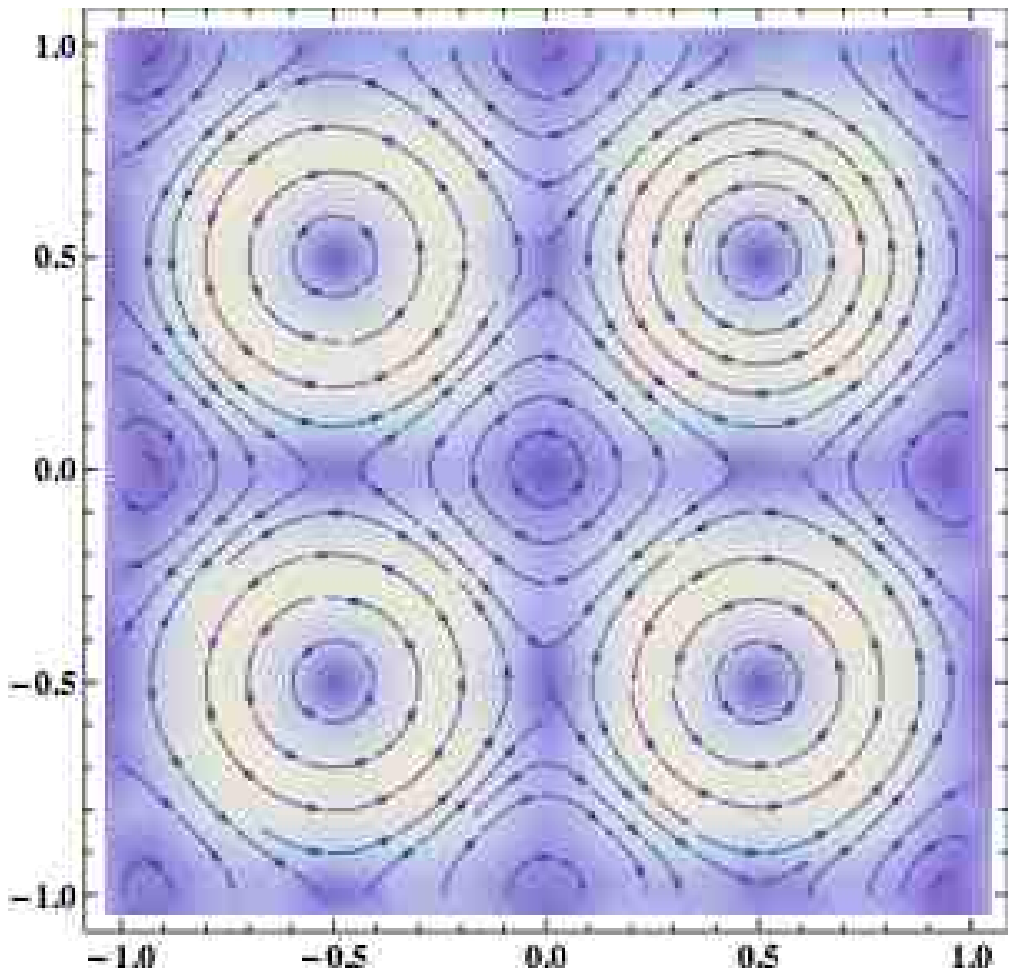}}
\subfigure[Director $\hat n(x)$ in a square vortex-disclination lattice]{\label{fig:currents-square-b}
\includegraphics[width=0.33\textwidth]{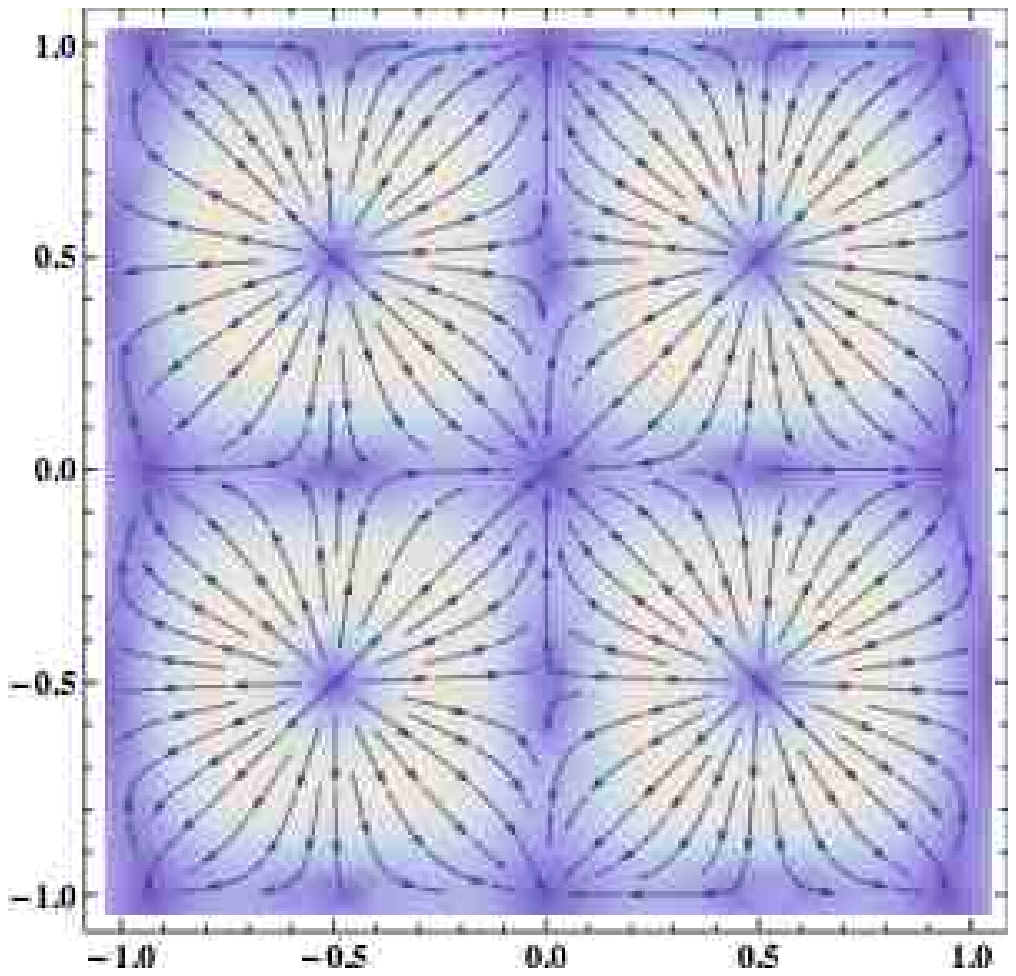}}
\end{center}
\caption{Phase structure of the SC and nematic order parameters in a vortex-disclination square lattice }
\label{fig:currents-square}
\end{figure}

\begin{figure}[hbt]
\begin{center}
\subfigure[Contours of constant SC order parameter, $|\chi_\triangle|^2=\mbox{constant}$ ]{\label{fig:currents-triangle-a}
\includegraphics[width=0.33\textwidth]{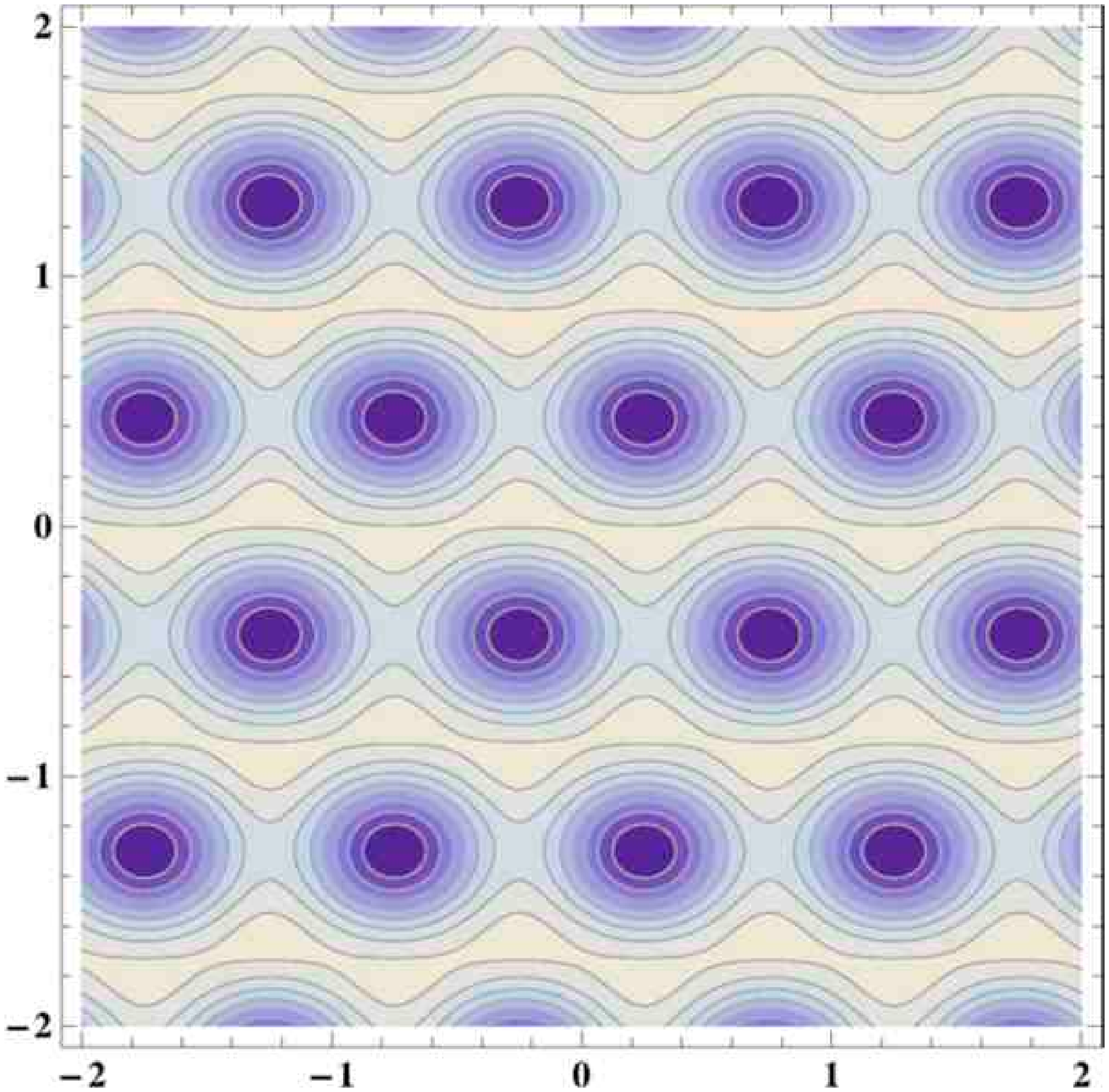}}
\subfigure[Contours of constant nematic order parameter, $|\vec\nabla|\chi_\triangle||=\mbox{constant}$ ]{\label{fig:currents-tiangle-b}
\includegraphics[width=0.33\textwidth]{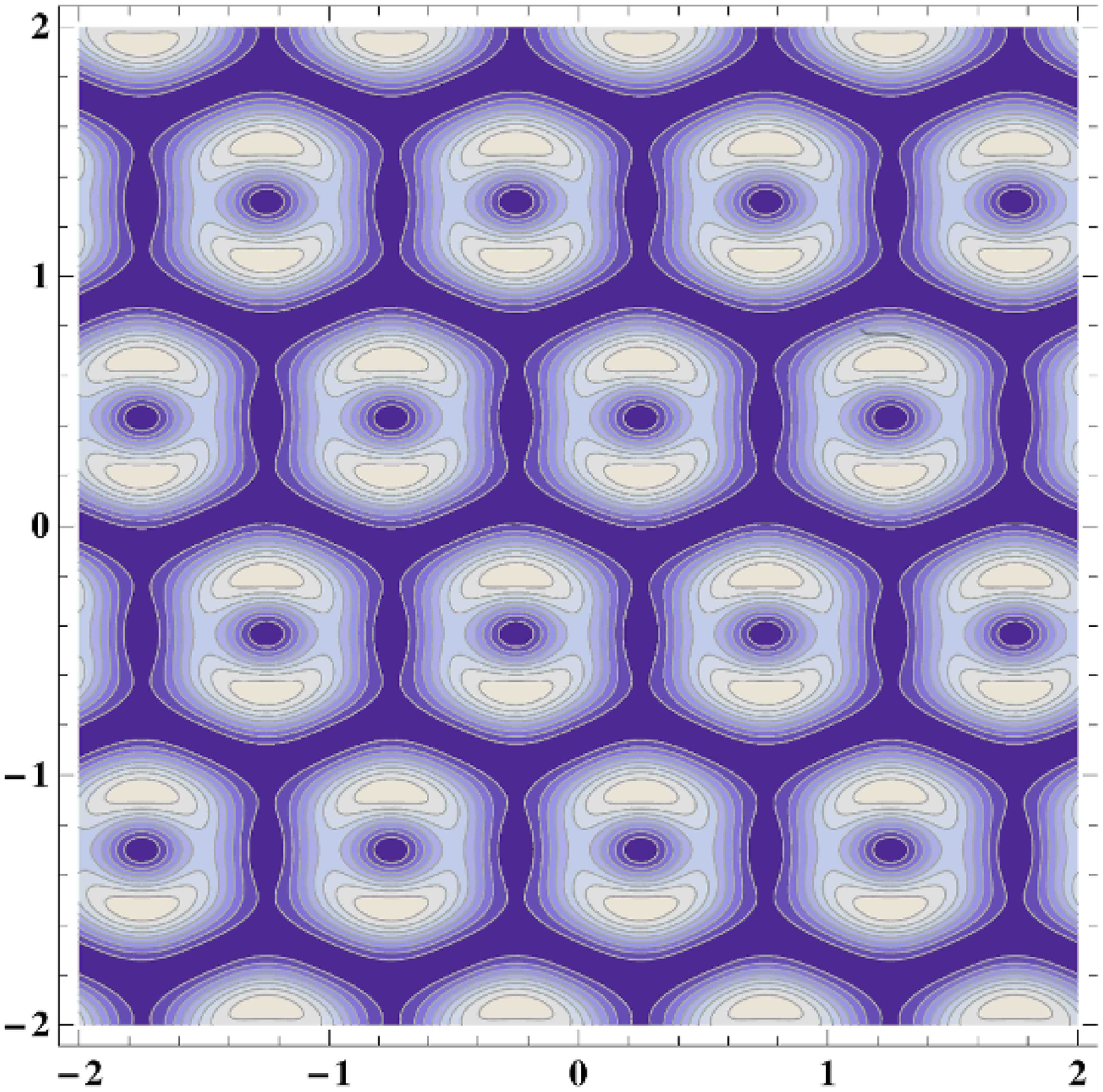}}
\end{center}
\caption{Vortex and Disclination lattice in a triangular (or hexagonal) configuration.}
\label{fig:VDTriangle}
\end{figure}

\begin{figure}[hbt]
\begin{center}
\subfigure[Currents in a triangular vortex-disclination lattice]{\label{fig:currents-triangular-a}
\includegraphics[width=0.33\textwidth]{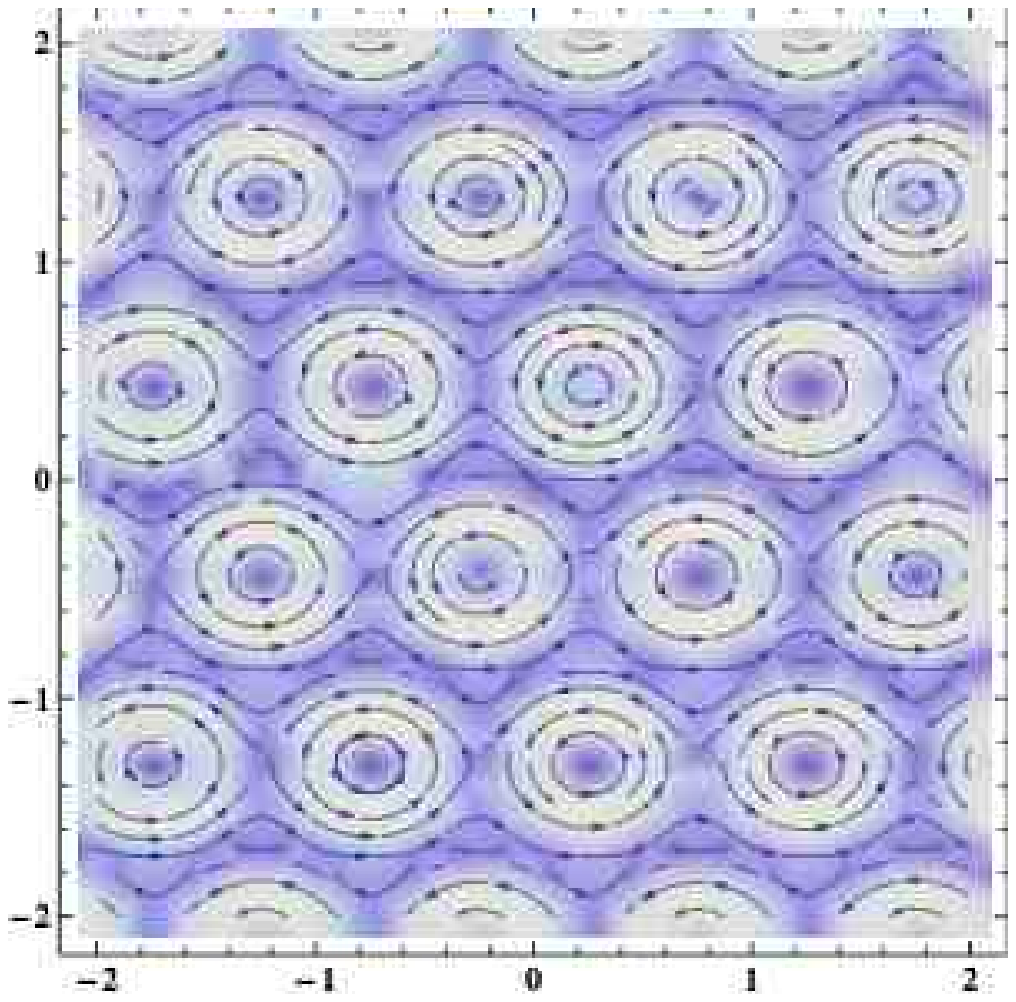}}
\subfigure[Director $\hat n(x)$ in a triangular vortex-disclination lattice]{\label{fig:currents-triangular-b}
\includegraphics[width=0.33\textwidth]{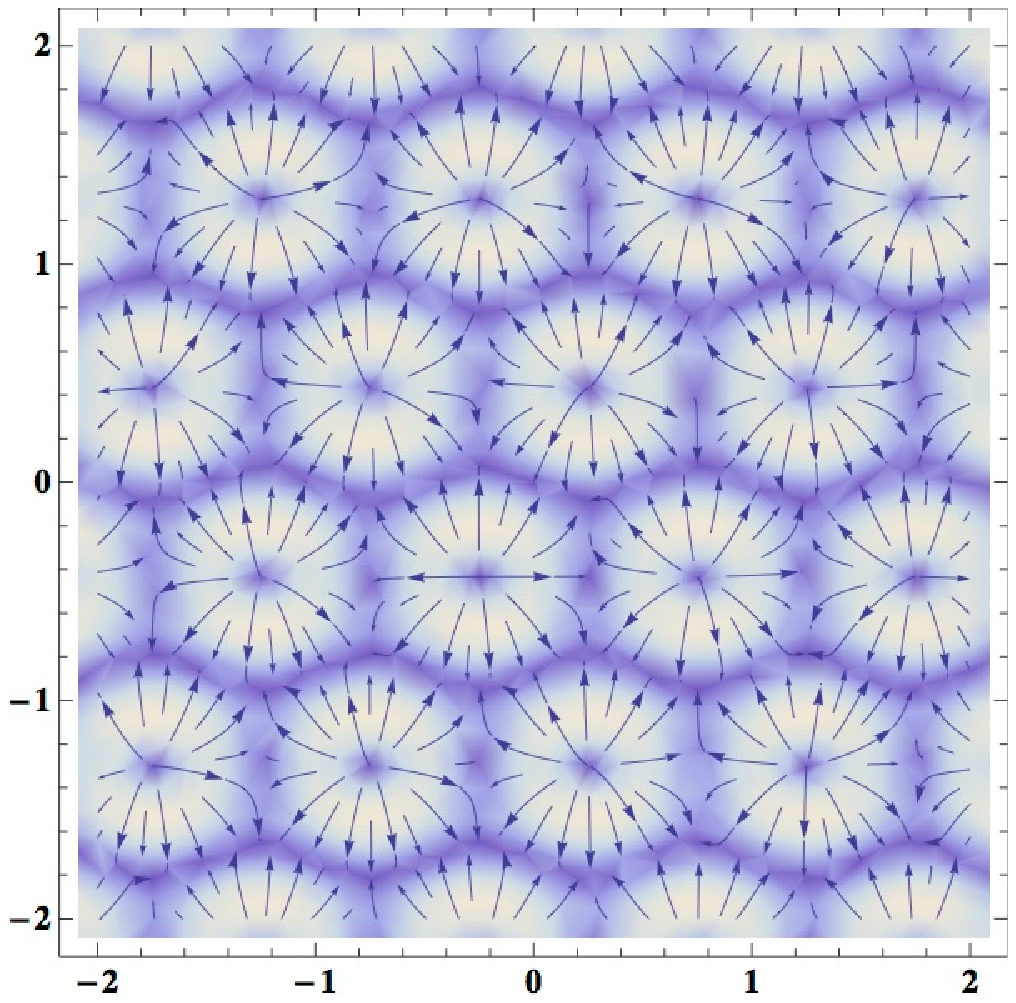}}
\end{center}
\caption{Structure of the streamlines of currents and the nematic director in a vortex-disclination triangular lattice }
\label{fig:currents-triangular}
\end{figure}

The nematic order parameter has a different structure than the superconductor one because  it does not  couple with the magnetic field in a minimal way. In order to determine it,  it is necessary to look at the geometrical coupling between both order parameters. Specifically, the second term of Eq. (\ref{eq:geometric}) displays the form, 
\begin{equation}
\Lambda\alpha_s S \rho^2 \left\{ \hat n\cdot \left(\vec \nabla\theta+4e \vec A\right)\right\}^2=\frac{\Lambda}{\alpha_s\rho^2}(\vec N\cdot \vec J^{sc})^2\; , 
\end{equation} 
where $\vec N=S^{1/2} \hat n$ is a vector with the same direction of  the director  and the supercurrent  $\vec J_{sc}=\alpha_s\rho^2(\vec\nabla\theta+4e \vec A)$. It is clear that, for $\Lambda>0$,  this term is minimum when the director is perpendicular to the supercurrent. This effect was deduced in the previous section for an isolated vortex with axial symmetry. The physical consequence is that vortices are necessarily tighten to disclinations in the nematic-superconductor state. The same physics applies here where we have no axial symmetry and a high density of vortices. The key observation to determine the nematic order parameter is that the streamlines of $\vec J_{sc}$ and the contours of constant $\rho$ coincide. To see this, we note that  the ground state satisfies, near $H_{c2}$, the first order equation
\begin{equation}
(D_x-iD_y)\psi =0  \;.
\label{eq:firstorder}
\end{equation}  
With this property, it is immediate to show that 
\begin{equation}
J_i^{sc}=-\alpha_s\epsilon_{ij} \nabla_j\rho^2\sim -\epsilon_{ij} \nabla_j|\chi|^2 \; .
\end{equation}
Thus, by choosing  
\begin{equation}
\vec N(x,y)=C_2 \vec\nabla|\chi(x,y)|^2 \; , 
\end{equation}
where $C_2$ is a variational parameter, we guarantee that locally $\vec N(\vec x)\cdot\vec J^{sc}(\vec x)=0$.  
In Fig. (\ref{fig:VDSquare}),  we illustrate the vortex and the disclination lattice profiles for the tetragonal symmetry case.  The vortex contours are drawn from the equation $|\chi_\Box|^2=\mbox{constant}$, while the disclination profile is computed from  
$|\vec\nabla|\chi_\Box|^2|=\mbox{constant}$. These pictures  represent the modulus of the order parameters. The phase structure is shown in Fig. (\ref{fig:currents-square}).  In Fig. (\ref{fig:currents-square-a}) we depict the supercurrent $\vec J_{sc}(\vec x)$, while in Fig.(\ref{fig:currents-square-b}) we show the director configuration $\vec N(\vec x)/|\vec N|$, locally perpendicular to the current.   
It is important to note that, while  the direction of the current determines the magnetization,  the direction of the director is meaningless, since the nematic order parameter is a quadratic function of the director. Thus, the particular configuration shown in  Fig. (\ref{fig:currents-square-b}), as well as  all the configurations obtained by  locally rotating  the director by  $\pi$, represent  exactly the same state. This is at the stem of the nematic symmetry. 
%%%%%%%%%%%%%%%%%%%%
%%%%%%%%%%%%%%%%%%%%
%%%%%%%%%%%%%%%%%%%%
%%%%%%%%%%%%%%%%%%%%
In Figs. (\ref{fig:VDTriangle}) and  (\ref{fig:currents-triangular}) we show the equivalent modulus and phase representation in the triangular lattice case.
%%%%%%%%%%%%%%%%%%%%
%%%%%%%%%%%%%%%%%%%%
%%%%%%%%%%%%%%%%%%%%
%%%%%%%%%%%%%%%%%%%%

\begin{figure}[hbt]
\begin{center}
\includegraphics[width=0.33\textwidth]{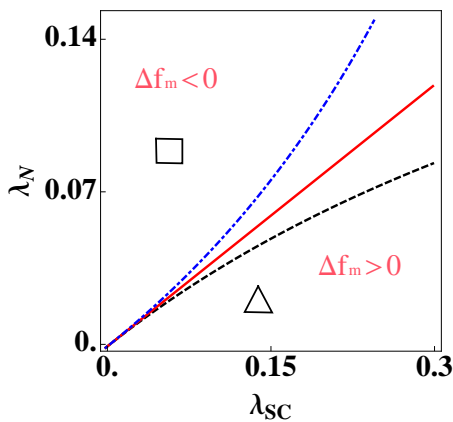}%}
\end{center}
\caption{$\Delta f_m=0$, continuous line $\tilde v=0$, dashed line $\tilde v=1$, dash-dot line $\tilde v=-1$ }
\label{fig:Deltafm}
\end{figure}

The next step is to  compute the free energy as a function of the variational parameters $C_1$ and $C_2$.  Near the transition, the relevant contribution comes from the potentials. The derivative terms are higher order corrections that do not change the qualitative results. 
The free energy density has  the following  form
\begin{equation}
f(C_1,C_2)=f_{SC}+f_{N}+f_{NSC} \; ,
\label{eq:fc1c2}
\end{equation}
 where the main three contributions are
\begin{eqnarray}
f_{SC}&=& a C_1^2\langle|\chi|^2\rangle+\frac{b}{2} C_1^4\langle|\chi|^4\rangle \; ,  \\
f_{N}&=& t C_2^4\left\langle\left(\vec\nabla|\chi|^2\right)^4\right\rangle+\frac{u}{2} C_2^8
\left\langle\left(\vec\nabla|\chi|^2\right)^8\right\rangle \; , \\
f_{NSC}&=& \frac{v}{2} C_1^2 C_2^4 
\left\langle |\chi|^2 \left(\vec\nabla|\chi|^2\right)^4\right\rangle \; .
\end{eqnarray}
We have defined the average 
$
\langle \ldots \rangle=(1/A)\int_A  d^2x\ldots
$
in which $A$ is the area of the sample.
Minimizing with respect to $C_1$ and $C_2$, 
\begin{equation}
\frac{\partial \left(f_{SC}+f_{NSC}\right)}{\partial C_1}=0\mbox{~~~~,~~~~}\frac{\left(\partial f_{N}+f_{NSC}\right)}{\partial C_2}=0 \; , 
\end{equation}
and computing the energy at this minimum we find, 
\begin{equation}
f_m=-\frac{a^2}{2b}\frac{1}{\beta_A}-\frac{t^2}{2u}\frac{1}{\beta_N}+\frac{a t v}{2bu}\frac{\beta_I}{\beta_A\beta_N}\; , 
\label{eq:fm}
\end{equation}
where the numerical coefficients $\beta_A$, $\beta_N$ and $\beta_I$  depend only on the lattice symmetry and are given by,  
\begin{eqnarray}
\beta_A&=&\frac{\langle|\chi|^4\rangle}{\langle|\chi|^2\rangle^2}\; , 
\label{eq:betaA}
\\
\beta_N&=&\frac{\left\langle\left(\vec\nabla|\chi|^2\cdot\vec\nabla|\chi|^2\right)^4\right\rangle}
{\left\langle\left(\vec\nabla|\chi|^2\cdot\vec\nabla|\chi|^2\right)^2\right\rangle^2} \; , 
\label{eq:betaN}
\\
\beta_I&=&\frac{\left\langle|\chi|^2\left(\vec\nabla|\chi|^2\cdot\vec\nabla|\chi|^2\right)^2\right\rangle}
{\left\langle |\chi|^2\right\rangle\left\langle\left(\vec\nabla|\chi|^2\cdot\vec\nabla|\chi|^2\right)^2\right\rangle}\; .
\label{eq:betaI}
\end{eqnarray}
Eq. (\ref{eq:betaA}) is the well known  Abrikosov coefficient\cite{Abrikosov-1957}. On the other hand, Eq. (\ref{eq:betaN}) defines an analog parameter for the nematic order and Eq. (\ref{eq:betaI}) takes into account correlations between the two order parameters. 
We have numerically computed these coefficients for the triangular as well as the square lattice (see appendix \ref{ap:beta}). The results are depicted in table \ref{tb:beta}. We are showing these coefficients with two decimal digits because this is sufficient for our purpose. However, we could compute them with any precision needed (see appendix \ref{ap:beta}). 
The first line of table \ref{tb:beta} depicts the known results for the Abrikosov coefficients for the triangular as well as for the square  lattice. Since $\beta_A^{\triangle}< \beta_A^\Box$, the  triangular lattice of vortices is more stable than the square lattice. Interestingly, we found that   $\beta_N^{\triangle}> \beta_N^\Box$, making more favorable the square lattice of disclinations. Then, there is a competition between vortices and disclinations and the form of the most stable configuration will depend on the parameters of the potentials. 
%%%%%%%%%%%%%%
\begin{table}
\begin{tabular}{|c|c|c|}
\hline
 Free energy parameters  & Triangular lattice & Square lattice\\
 \hline
$\beta_A$ &  1.16 & 1.18 \\
\hline
$\beta_N$ &  2.93  & 2.53\\
\hline
$\beta_I$ &  1.60 &  1.21\\
\hline
\end{tabular}
\caption{Free energy coefficients for triangular and square lattices. $\beta_A$ is the known Abrikosov coefficient. $\beta_N$ is an analog coefficient for the nematic phase, given by Eq. (\ref{eq:betaN}). $\beta_I$ describes correlation contributions given by Eq. (\ref{eq:betaI}). }
\label{tb:beta}
\end{table}
%%%%%%%%%%%%%%%%%%
To see this more clearly, let us compute the energy difference between the triangular and the square lattice of vortices attached to disclinations. Using Eq. (\ref{eq:fm}) with the values of $\beta_A$, $\beta_N$ and $\beta_I$ taken from table \ref{tb:beta} we find, 
\begin{eqnarray}
\Delta f_m&=& f_m^\Box-f_m^\triangle \nonumber  \\
&=& 0.02 \lambda_{SC}-0.05 \lambda_N-0.07 \tilde v \lambda_{SC}\lambda_N  \; ,
\label{eq:Deltafm}
\end{eqnarray}
where $\lambda_{SC}=a^2/2b$ and $\lambda_{N}=t^2/2u$ are the superconductor and the nematic  condensation energy respectively,   and we have renormalized the coupling constant $\tilde v=v/a t$. We can clearly see a competition between the superconductor and the nematic contribution. The first term of Eq. (\ref{eq:Deltafm}), coming from the superconductor free energy, favors the triangular lattices configuration. On the other hand, the second term, coming from the nematic free energy, favors the square lattice configuration. The interaction contribution depends on the sign of $\tilde v$. Positive couplings $v>0$ strengths the square lattice configuration, while negative couplings $v<0$ favor the triangular one.  We have depicted this competition in Fig. (\ref{fig:Deltafm}), where we show 
the line $\Delta f_m=0$ for three different values of the coupling $\tilde v=-1,0,1$. 
%%%%%%%%%%%%%%%%%%%%
%%%%%%%%%%%%%%%%%%%%
In the region $\Delta f_m>0$ the system tends to form a triangular lattice of vortices attached to disclinations, while for  $\Delta f_m<0$, the state is arranged in a square lattice configuration. Thus, the curves $\Delta f_m=0$ represent a structural phase transition between these two different discrete symmetries.

%%%%%%%%%%%%%%%%%%%%%%%%%%%%%%%%%%%%%%%%%
\section{\label{sec:discussion} Conclusion and discussion}
%%%%%%%%%%%%%%%%%%%%%%%%%%%%%%%%%%%%%%
%%%%%%%%%%%%%%%%%%%%
The charge-$4e$ nematic superconductor is an homogeneous state of electronic matter that breaks gauge as well as rotational symmetry. It can be understood as a condensation of  four particles of charge $e$ or, equivalently, as a melted state of  pair density waves, obtained by the proliferation of double dislocations\cite{berg-2009}. The $4e$-NSC state has essentially two types of topological excitations: half-vortices, or vortices with half a quantum flux, and disclinations. In this paper, we have analyzed in detail the structure of these topological defects in different regimes of magnetic fields. 

We have built up a Ginzburg-Landau theory for the simplest superconductor order parameter coupled with a two-dimensional nematic order. The SC order parameter is a complex function while the nematic one is a  symmetric traceless tensor of order two.    The main effect of local nematicity is to induce a  deformation of the metric, in such a way that the SC order parameter ``feels''  an effective curved space. As a result, the nematic director has a tendency to be perpendicular or parallel to the supercurrent, depending on the sign of the geometrical coupling. Thus, vortices induce disclinations.
We have minimized the Ginzburg-Landau energy in two regimes: for magnetic fields near $H_{c1}$ where the vortices are extremely diluted and near $H_{c2}$ where the system develops a high density of vortices.

Computing the energy of a vortex-disclination configuration, we obtained an attractive force as a function of the distance $R$  between the cores of the vortex and the disclination,  that does not depend on the sign of the topological index.  At short distances, the potential is harmonic, $V(R)\sim \Omega^2 R^2$, where the typical frequency depends on the geometrical coupling constant and the SC condensation energy. At large distances,     
the potential remains attractive and it  is logarithmic $V(r)\sim \ln (R/{\cal R}_v)$, where ${\cal R}_v$ is the vortex core radius. Interestingly, ${\cal R}_v$ is a decreasing function of the geometrical coupling constant while the core of the disclination is very weakly dependent. 

Increasing the external magnetic field, we reach a regime of high density of vortices, where each vortex is tightly bounded to a disclination. In this high density regime, the director has also a strong tendency to be perpendicular to the supercurrent. We explored the possibility of forming vortex/disclination lattices.  We have implemented a variational calculation, analog to the Abrikosov lattice, but taking into account the effect of disclinations. Comparing the  free energy for different configurations, we found that, while the vortices contribution is minimum for triangular symmetry, the disclination contribution is minimized by a square lattice. Then, there is a competition that, depending on the SC and nematic condensation energies, produces a structural phase transition between lattices with different symmetries. 

Interactions between vortices and disclinations should have profound influence in the 
elastic response of the material. 
On the one hand, vortices induce strain and consequently, the energy of the vortex lattice has a contribution from the vortex-induced strains\cite{Cano-2003}. On the other hand, disclinations induce torque\cite{deGPr1998} and there should be a corresponding contribution to the disclination lattice.  
Moreover, since the strain and the nematic order parameters are second order tensor, they should also be coupled. 
In this way the magneto-elastic properties of the vortex-disclination structure should be non-trivial.
We believe that the magneto-elastic properties should contain signatures that, in principle, could allow to study the  4e-NSC state by means of  experimental magneto-mechanical probes, such as the application of strain\cite{Locquet-1998} or acoustic waves\cite{Sonin-1996}.
 
\acknowledgments
We are in debt with  Eduardo Fradkin for very  useful discussions.
The Brazilian agencies {\em Conselho Nacional de Desenvolvimento Cient\'\i fico e Tecnol\'ogico} (CNPq), {\em Funda\c c\~ao  Carlos Chagas Filho de Amparo \`a Pesquisa do Estado do Rio de Janeiro} (FAPERJ), and {\em Coordena\c c\~ao de Aperfei\c coamento de Pessoal de N\'\i vel Superior} (CAPES) are acknowledged for partial financial support. NJL is financed by a MSc fellowship by CAPES.  RVC is financed by a doctoral fellowship by FAPERJ.  D.G.B also acknowledges partial financial support by the  Associate Program of the Abdus Salam International Centre for Theoretical Physics, ICTP, Trieste, Italy.

\appendix

%%%%%%%%%%%%%%%%%%%%%%%
\section{Variational Ansatz}
\label{ap:Ansatz}

Although Eqs. (\ref{eq:vortex-ansatz}) and (\ref{eq:disclination-ansatz}) are very good and simple Ansatz for the vortex and the disclination profile, it is possible to perform a systematic  procedure to improve it, getting closer to the exact solutions. For this, we define the function 
\begin{equation}
\tau(x)=\frac{x}{\sqrt{1+x^2}}\; .
\end{equation}
The simple Ansatz of Eqs. (\ref{eq:vortex-ansatz}) and (\ref{eq:disclination-ansatz}) now read, 
\begin{eqnarray}
f(r)&=&\tau(\sqrt{2}\kappa_f r)  \\
g(r)&=&\tau^2(\kappa_g r) .
\end{eqnarray}
The function $\tau(r)$ maps the open domain $r\epsilon [0,\infty)$ to the compact interval $\tau\epsilon [0,1]$. Then, we  write a complete solution in the form 
\begin{eqnarray}
f(r)&=&\tau(\sqrt{2}\kappa_{f}r)+\Gamma_f[\tau(\sqrt{2}\kappa_{f}r)] \; , \\
g(r)&=&\tau^2(\kappa_{g}r)+ 
 \tau(\kappa_{g}r)  
 \Gamma_g[\tau(\kappa_{g}r)]  \; , 
\end{eqnarray}
where the arbitrary functions $\Gamma_{f,g}[\tau]$ satisfy $\Gamma_{f,g}[0]=\Gamma_{f,g}[1]=0$. 
These boundary conditions allow us to represent  $\Gamma_{f,g}[\tau]$ by means of  $\sin$-Fourier series\cite{Oxman-2016}. 
Based on that, we propose the following  Ansatz for the full vortex-disclination solution:
\begin{eqnarray}
f(r)&=&\tau(\sqrt{2}\kappa_{f}r)+\sum_{n=1}^{\infty} a_n \sin\left[n\pi \tau(\sqrt{2}\kappa_{f}r)\right] \; ,
\label{eq:ansatzf} \\
g(r)&=&\tau^2(\kappa_{g}r)+ 
 \tau(\kappa_{g}r) \sum_{n=1}^{\infty} b_n \sin\left[n\pi \tau(\kappa_{g}r)\right] \; .
 \label{eq:ansatzg}
\end{eqnarray}
 This Ansatz has the correct behavior at the boundaries $r\sim 0$ ($\tau\sim 0$) and $r\sim \infty$ ($\tau\sim 1$) and it is completely determined by the set of Fourier coefficients   
$\{a_1,a_2,\ldots\}$ and $\{b_1,b_2,\ldots\}$. The Fourier coefficients  can be computed in two ways. We can expand Eqs. (\ref{eq:ansatzf}) and  (\ref{eq:ansatzg}) in Taylor series  for small $r$ and compare the coefficients to  the ones computed in Eqs. (\ref{eq:rhoseries}) and (\ref{eq:Sseries}).  An alternative procedure is to plug the Ansatz into  the free energy and minimize it with respect to a finite set of Fourier coefficients. Amazingly, we found that the Fourier series converge very fast. Indeed, after the second harmonic, it is no longer possible to distinguish any significant difference within the graphic precision. This is so because the leading order $f\sim \tau$ and $g\sim \tau^2$ is an excellent qualitative description and it is very ``near" (in functional space) to  the exact solution.  

%%%%%%%%%%%%%%%%%%%%%%%%%%%%%%%
\section{Vortex-Disclination interaction energy}
\label{ap:VD}
Considering the vortex and the disclination configuration given by  Eqs. (\ref{eq:vortex}) and 
(\ref{eq:disclination}) and shown in Fig. (\ref{fig:VDR}) we can compute, 
\begin{equation}
\left(\hat n\cdot \vec\nabla\theta\right)^2=\frac{R^2\sin^2\varphi}{r^2\left(r^2+R^2-2rR\cos\varphi\right)} \; , 
\end{equation} 
where $(r,\varphi)$ are usual polar coordinates given by $x=r\cos\varphi$ and $y=r\sin\varphi$. Then, the energy $\Delta F(R)=F(R)-F(0)$  is given by 
\begin{eqnarray}
\Delta F(R)&=&\lambda\int d^2 x\; \left(\mathbf{n}(x)\cdot \mathbf{\nabla}\theta(x,R)\right)^2
 \label{ap:DeltaF}\\
&=&\lambda\Lambda\int_a^L \frac{dr}{r}\int_0^{2\pi} d\varphi
\frac{R^2\sin^2\varphi}{r^2+R^2-2rR\cos\varphi}   \; , 
\nonumber
\end{eqnarray}
where $a$ is the vortex core and $L$ is the linear size of the sample.
Introducing the dimensionless variable $z=r/R$ we immediately find, 
\begin{equation}
\Delta F(R)=\pi\lambda\int_{a/R}^{L/R} \frac{dz}{z} I_{\varphi}(z) \; , 
\label{eq:intz}
\end{equation}
where we have defined
\begin{eqnarray}
I_{\varphi}(z)&=&\frac{1}{\pi}\int_0^{2\pi} d\varphi \frac{\sin^2\varphi}{z^2+1-2z\cos\varphi} \nonumber \\
&=&\frac{1}{2z^2}\left\{1+z^2-(1+z)|1-z|\right\}  \; .
\end{eqnarray}
Thus, 
\begin{equation}
I_{\varphi}(z)=\left\{ 
\begin{array}{lll}
1 & ~ & z<1 \\
\frac{1}{z^2} & ~ & z\geq 1
\end{array}
\right.   \; .
\label{eq:Iphi}
\end{equation}
Introducing Eq. (\ref{eq:Iphi}) into Eq. (\ref{eq:intz})
we have
\begin{equation}
\Delta F(R)=\pi \lambda \left\{\int_{a/R}^1\frac{dz}{z}+\int_1^{L/R}\frac{dz}{z^3}\right\} \; .
\label{eq:int2}
\end{equation}
We see that the first term of Eq. (\ref{eq:int2}) has  a logarithmic divergence regulated by the vortex core  $a$. This divergence will dominate the interaction energy. The integrals in Eq. (\ref{eq:int2}) can be done without any difficulties obtaining
\begin{equation}
\Delta F(R)=\lambda\pi\left\{\ln\left(\frac{R}{a}\right) -\frac{1}{2}\left(\frac{R}{L}\right)^2+\frac{1}{2}\right\} \; .
\end{equation}
Considering $a<<R<<L$, we can take the thermodynamic limit $L\to \infty$ and ignore the unimportant constant contribution obtaining
\begin{equation}
\Delta F(R)=\lambda\pi\ln\left(\frac{R}{a}\right) \; .
\end{equation}
that coincides with Eq. (\ref{eq:Deltalog}).
Eqs. (\ref{ap:DeltaF}) is invariant under the transformation $\lambda\to -\lambda$, $\varphi\to\varphi + \pi /2$. For this reason $\Delta F(R)$ does not depend on the sign of $\lambda$.

%%%%%%%%%%%%%%%%%
\section{$\beta$ coefficients}
\label{ap:beta}
In this appendix we sketch the explicit calculation of the coefficients $\beta$, displayed in table (\ref{tb:beta}). 

\subsection{$\beta_A$}
Let us review the computation of the well known Abrikosov parameter $\beta_A$\cite{Abrikosov-1957,Kleiner-1964,Brandt-1995}. We want to compute 
\begin{equation}
\beta_A=\frac{\langle|\chi|^4\rangle}{\langle|\chi|^2\rangle^2} \; , 
\label{eqAp:betaA}
\end{equation}
for different lattice symmetries. The main observation is that $|\chi(x,y)|^2$ is periodic,
\begin{equation}
|\chi(\vec r+\vec R_{n,m})|^2=|\chi(\vec r)|^2 \; , 
\end{equation} 
where $\vec R_{n,m}=(m x_1+n x_2; n y_2)$, with $n,m=0,\pm 1,\pm 2,\ldots$, are lattice vectors. In the square lattice case, 
$x_1=y_2=a$, $x_2=0$, where $a$ is the lattice constant.
For triangular symmetry, $x_1=a$, $x_2=x_1/2$ and $y_2=x_1 \sqrt{3}/2$.
Vectors in the reciprocal lattice are written as $K_{n,m}=(2\pi/x_1y_2)(m y_2; -m x_2+n x_1)$ in such a way that $\vec R_{n,m}\cdot \vec K_{n,m}= (n^2+m^2)2\pi$.
Thus, it is possible to represent the order parameter  as a Fourier series of the form, 
\begin{equation}
|\chi(x,y)|^2=\sum_{n,m} a_{n,m}\;   e^{i \vec K_{n,m}\cdot \vec r}\; , 
\label{eqAp:Fourier}
\end{equation}
where $a^*_{n,m}=a_{-n,-m}$.
The Fourier coefficients are computed by  inverting this equation and using  Eqs. (\ref{eq:chibox}) and (\ref{eq:chitriangle}). We find, 
\begin{equation}
a_{n,m}=(-1)^{nm} e^{-\frac{1}{8\pi} |\vec K_{n,m}|^2} \; .
\label{eqAp:Fcoefficients}
\end{equation}

The Abrikosov coefficient can be cast in terms of the Fourier coefficients by replacing Eq. (\ref{eqAp:Fourier}) into Eq. (\ref{eqAp:betaA}) and performing the integrals, 
\begin{equation}
\beta_A=\sum_{n,m} a_{n,m}^2\; .
\end{equation}
Then, using Eq. (\ref{eqAp:Fcoefficients}) we immediately find
\begin{equation}
\beta_A=\sum_{n,m} e^{-\frac{1}{4\pi} |\vec K_{n,m}|^2}\;.
\end{equation}
Computing this expression explicitly for different geometries we have
\begin{equation}
\beta_A=\sum_{n,m}  e^{-\pi(n^2+m^2)}\sim 1.18 \; .
\end{equation}
for the square lattice and 
\begin{equation}
\beta_A=\sum_{n,m}  e^{-\frac{2\pi}{\sqrt{3}}(n^2+m^2-nm)} \sim 1.16 \; .
\end{equation}
for the triangular one. Notice that, although $\beta_A$ is  given by a series,  it converges exponentially. Thus, the first few terms give an excellent approximation to the numerical value.  

\subsection{$\beta_N$}
The computation of the nematic coefficient
\begin{equation}
\beta_N=\frac{\left\langle\left(\vec\nabla|\chi|^2\cdot\vec\nabla|\chi|^2\right)^4\right\rangle}
{\left\langle\left(\vec\nabla|\chi|^2\cdot\vec\nabla|\chi|^2\right)^2\right\rangle^2}
\label{eqAp:betaN}
\end{equation}
follows exactly the same lines that the computation of the Abrikosov coefficient.  There are essentially two differences. 
It contains more powers of the order parameter and  it depends on its derivatives.  The main object that enter the computation of $\beta_N$ is 
\begin{eqnarray}
\lefteqn{
\vec\nabla|\chi|^2\cdot \vec\nabla|\chi|^2=} \label{eqAp:nablachi}  \\
&-&\sum_{n,m}\sum_{p,q} a_{n,m}a_{p,q}\vec K_{n,m}\cdot \vec K_{p,q}\;   e^{i (\vec K_{n,m}+\vec K_{p,q})\cdot \vec r}
\; ,  \nonumber 
\end{eqnarray}
where the Fourier coefficients are given by Eq. (\ref{eqAp:Fcoefficients}). Thus, the numerator of Eq. (\ref{eqAp:betaN}), using Eq. (\ref{eqAp:nablachi}) and performing the integrals is 
\begin{eqnarray}
\lefteqn{
\left\langle\left(\vec\nabla|\chi|^2\cdot\vec\nabla|\chi|^2\right)^4\right\rangle= \sum_{n_1,m_1}\!\!\ldots\sum_{n_4,m_4}\sum_{p_1,q_1}\!\!\ldots\sum_{p_4,q_4}
 } \nonumber \\
&\times&\delta\left(\sum_\ell (n_\ell+p_\ell)\right)\delta\left(\sum_\ell (m_\ell+q_\ell)\right) \times
\\
&\times&\prod_{i=1}^4 e^{-\frac{1}{8\pi} \left(|\vec K_{n_i,m_i}|^2+|\vec K_{p_i,q_i}|^2\right)}\left(\vec K_{n_i,m_i}\cdot \vec K_{p_i,q_i}\right) \; .
\nonumber 
\end{eqnarray}
In this expression all the series converge exponentially. For this reason, the numerical computation is not difficult, since very few terms gives a reasonable approximation.  By computing these sums explicitly for the triangular and the square lattice we found  $\beta_N= 2.93$   and  $\beta_N=2.53$  respectively,  as shown in table (\ref{tb:beta}). 

The computation of $\beta_I$  follows exactly the same lines as $\beta_N$ without additional difficulties.

%\bibliography{nematic-melting,nematics-D311016}

%merlin.mbs apsrev4-1.bst 2010-07-25 4.21a (PWD, AO, DPC) hacked
%Control: key (0)
%Control: author (8) initials jnrlst
%Control: editor formatted (1) identically to author
%Control: production of article title (-1) disabled
%Control: page (0) single
%Control: year (1) truncated
%Control: production of eprint (0) enabled
%

\end{document}